\documentclass[preprint,12pt]{aastex}

\shorttitle{Tidal Stream Fitting}
\shortauthors{Cole et al.}

\begin{document}

\title{Maximum Likelihood Fitting of Tidal Streams With Application to the Sagittarius Dwarf Tidal Tails}

\author{
Nathan Cole,\altaffilmark{\ref{RPI_A}}
Heidi Jo Newberg,\altaffilmark{\ref{RPI_A}}
Malik Magdon-Ismail,\altaffilmark{\ref{RPI_CS}}
Travis Desell,\altaffilmark{\ref{RPI_CS}}
Kristopher Dawsey,\altaffilmark{\ref{RPI_A}}
Warren Hayashi,\altaffilmark{\ref{RPI_A}}
Xinyang (Fred) Liu,\altaffilmark{\ref{RPI_A}}
Jonathan Purnell,\altaffilmark{\ref{RPI_CS}}
Boleslaw Szymanski,\altaffilmark{\ref{RPI_CS}}
Carlos Varela,\altaffilmark{\ref{RPI_CS}}
Benjamin Willett,\altaffilmark{\ref{RPI_A}}
James Wisniewski\altaffilmark{\ref{RPI_A}}
}

\altaffiltext{1}{Dept. of Physics, Applied Physics and Astronomy, Rensselaer
Polytechnic Institute Troy, NY 12180; astro@cs.rpi.edu\label{RPI_A}}

\altaffiltext{2}{Dept. of Computer Science, Rensselaer
Polytechnic Institute Troy, NY 12180; astro@cs.rpi.edu\label{RPI_CS}}

\begin{abstract}
We present a maximum likelihood method for determining the spatial properties of tidal debris and of the Galactic
spheroid.  With this method we characterize Sagittarius debris using stars with the colors of blue F turnoff stars in 
SDSS stripe 82.  The debris is located at 
($\alpha, \delta, R) = (31.37^{\circ}\pm0.26^{\circ}, 0.0^{\circ}, 29.22\pm 0.20$ kpc), with a (spatial)
direction given by the unit vector $<-0.991\pm0.007 {\rm ~kpc}, 0.042\pm0.033 {\rm ~kpc}, 0.127\pm0.046 {\rm ~kpc}>,$
in Galactocentric Cartesian coordinates, 
and with FWHM = $6.74\pm0.06$ kpc.  This $2.5^\circ$-wide stripe contains 0.9\% as many F turnoff stars as the current
Sagittarius dwarf galaxy.  Over small spatial extent, the debris is modeled as a cylinder with 
a density that falls off as a Gaussian with distance from the axis, while the smooth component of the 
spheroid is modeled with a Hernquist profile.  We assume that the absolute magnitude of F turnoff stars 
is distributed as a Gaussian, which is an improvement over previous methods which fixed the absolute 
magnitude at $\bar{M}_{g_0} = 4.2$.  The effectiveness and correctness of the algorithm is demonstrated on a simulated set 
of F turnoff stars created to mimic SDSS stripe 82 data, which shows that we have a much greater accuracy
than previous studies.  Our algorithm can be applied to divide the stellar data into two catalogs: one which fits 
the stream density profile and one with the characteristics of the spheroid.  This allows us to effectively separate 
tidal debris from the spheroid population, both facilitating the study of the tidal stream dynamics and providing a 
test of whether a smooth spheroidal population exists. 
\end{abstract}


\section{Introduction\label{intro}}

\subsection{The Milky Way's Spheroid}
In the late 1980s, the Milky Way's spheroid population was described as slowly rotating, with density 
distribution given by $\rho \propto r^{-3.5}$ \citep{f87}.  There was a controversy concerning the flattening of the 
spheroid, since kinematic studies of the spheroid stars suggested a flatter spheroid \citep{gwk89} and star 
counts often suggested a more spherical spheroid \citep{b86}.  By early in the 21$^{\rm st}$ century, the spheroid was 
still thought of as a smooth power law distribution, but studies were
starting to show that the shape of the spheroid depended on the type of star being observed, and it was noted that
at least some of the spheroid, if not all, was composed of debris from hierarchical structure formation \citep{fb02}.
The Sagittarius (Sgr) dwarf galaxy \citep{igi94} and its associated tidal stream \citep{ynetal00, ietal01, ietal01-2} 
was the one known example of merging in the present day.
In the last ten years, the discovery of substructure has dominated the discussion of the Galactic spheroid.  The
discovery of substructure has been driven primarily by the Sloan Digital Sky Survey (SDSS) and the related
Sloan Extension for Galactic Understanding and Exploration (SEGUE), but other surveys such as the Quasar Equatorial
Survey Team (QUEST) and the Two Micron All Sky Survey (2MASS) have also been influential.  Discoveries include new
tidal debris streams, dwarf galaxies, and globular clusters.  

The positions, velocities, and metallicities of Sgr dwarf spheroidal tidal debris stars have been measured all 
across the sky \citep{nyetal02, betal03, mswo03, netal03, metal04, mdetal04, fellhauer, cetal07}.  These measurements 
have then been compared with models of tidal disruption \citep{jetal95, ietal97, il98, gfetal99, hw01, jetal02,law04,law05}.  Given that the data 
spans three spatial dimensions plus radial velocities, but both the models and data are presented in papers primarily
as two dimensional plots, it has been difficult to make detailed comparisons in multidimensional space.  
In this paper we develop a technique that would allow us to create a catalog of stars that has the same
spatial distribution as the stars in the Sgr tidal stream.  That catalog could then be made available to the modeling
community, who would be able to transform the coordinates into whatever system they find most natural to make
comparisons.

At least three newly discovered tidal debris streams are thought to be associated with dwarf galaxies, including:
the Monoceros stream in the Galactic plane \citep{nyetal02,ynetal03}, the Virgo Stellar Stream 
\citep{vetal01,duffau,nyetal07}, and the Orphan Stream (Grillmair 2006) in the Field of Streams\citep{fieldofstreams}.  
An additional piece of tidal debris was found in Triangulum-Andromeda by \citet{rmscp04} which might be part of the
Monoceros stream \citep{pmrgmnbyzg05}.
Tidal tails spanning many tens of degrees across the sky have been found around Pal 5 
\citep{oetal01,retal02,gd06Pal5}, and NGC 5466 \citep{gj06,beihw06}.  \citet{gd06} find a similar-looking 
tidal stream for which there is no known progenitor globular cluster.
Eight new low surface brightness dwarf galaxy satellites of the Milky Way have been discovered \citep{wetal05,
zetal06a,bootes, zetal06b,grillmair06, betal07cats, ietal07},
nearly doubling the number of known Milky Way dwarfs.  
\citet{betal07b} suggest that there is a new structural component of the spheroid called the Hercules-Aquila cloud.

With all of this substructure detected in the densities of spheroid stars, one wonders how much of the spheroid is smooth,
and what the shape of the smooth spheroid might be.  
It is no wonder that standard Galactic models, such as the the Besan\c{c}on model \citep{retal03}, which uses a power law density profile, do not match the SDSS star counts.  
The spheroid shape is a current controversy as \citet{h04} has found prolate models favorable while \citet{jetal05} has found oblate models best.  
The large number of very significant over-densities of 
stars in the spheroid show that pencil beam studies and studies that extrapolate from the solar neighborhood could 
be contaminated by local structure and may not be indicative of the global shape of the Milky Way 
spheroid.  Recent results from F turnoff stars from the SDSS suggest the smooth component of the spheroid might 
be asymmetric about the Galactic center, so that no axially symmetric spheroid model can be 
fit \citep{ny05,ny06,snfg06,xdh06}.  However, accurate measurements of the smooth component require that 
spatial substructure be identified and quantified --- otherwise it is difficult to know which stars to fit to 
smooth spheroid models.

We need to develop a model that is as complex as the data.  In this paper we start this process by developing a maximum likelihood method that can be used to fit a smooth stellar spheroid containing one
debris stream, using one 2.5$^\circ$ stripe of SDSS data, and show that our algorithm produces the correct results on a 
simulated dataset.  
In the future, we will extend the algorithm to include larger datasets with 
multiple pieces of debris.  The long-term goal is to develop a model of the spheroid that fits the observed, lumpy 
density distribution of stars in the spheroid.

\subsection{The Sloan Digital Sky Survey\label{sdss}}

The SDSS is a large, international collaboration that was originally established to find the largest structures of galaxies
in the Universe from an imaging survey of 10,000 sq. deg. of sky and spectra of 1,000,000 galaxies selected from the
photometry (from images).  
The SDSS can simultaneously obtain 640 spectra per pointing, using two 320-fiber double (blue/red) spectrographs.
The angular position of a galaxy in the sky is easy to measure from the imaging survey, and for
distant galaxies the distance is well estimated from a spectroscopic measurement of the Hubble redshift.  By combining the
imaging with the spectroscopic survey, the SDSS has built up a three dimensional picture of the distribution of galaxies
in the Universe using a dedicated 2.5 meter telescope at Apache Point Observatory in New Mexico \citep{yetal00}.  

Images are obtained with an array of thirty $2048 \times 2048$ pixel CCD cameras operated in a drift-scan 
(time delay and integrate, TDI) mode, which produces six ``scan lines" of imaging data that are each 13.6' wide and
grow longer at a rate of 15 degrees per hour for the length of that ``run."  In each of the six 
``scan lines," the sky is imaged in five optical
filters: $u, g, r, i,$ and $z$; the time at which each astronomical object is imaged is a few minutes different for 
each passband.  When a second set of six scan lines is observed that fill in the gaps between the first set of six
scan lines, they produce a contiguous ``stripe" of data of width $2.5^\circ$ and length that depends on the
length of time the sky was observed in the ``runs."  Each $2.5^\circ$-wide stripe follows a defined great 
circle on the sky, and is built up of two
or more ``runs" of data, as many runs as are needed to traverse the SDSS survey area.

The entire sky is divided up into 144 numbered stripes that start and end at the survey poles: 
$(l,b)=(209.33^\circ,-7^\circ)$ and $(l,b)=(29.33^\circ, 7^\circ)$.  
Stripes 10 and 82 are centered on the Celestial Equator, with stripe 10 in the North Galactic Cap and 
stripe 82 in the South Galactic Cap.  The other stripes are sequentially numbered with inclinations 2.5 degrees apart.
If the entire length ($180^\circ$) of every stripe were imaged, we would have $64,800$ sq. degrees of imaging data
for a sky that is $41,253$ sq. deg.  The overlaps between stripes increase towards the survey poles.  Since the 
SDSS generally observes only parts of the sky
that are more than $30^\circ$ from the survey poles, there is about 50\% overlap at the ends of the stripes and
very little overlap on the survey equator, which is at RA$=185^\circ$.  With the release of SDSS DR6 \citep{DR6}, a contiguous
8500 sq. deg. area of the North Galactic Cap is now publicly available.  It is pieced together with 31 adjacent SDSS
stripes.  Three stripes of data are available in the South Galactic Cap, including stripe 82 on the
Celestial Equator.    
Further information about the SDSS, including survey geometry, can be found in:
\citet{setal01}, \citet{aetal03}, \citet{figdss96}, 
\citet{getal98}, \citet{hsfg01}, \citet{pmhhkli03}, \citet{setal02}.

Because the imaging survey is large and well calibrated, and because the spectroscopic survey included many Galactic
stars, in some cases by design and in some cases accidentally, the SDSS has made a significant contribution to our
knowledge about the Milky Way, and in particular the discovery of substructure in the spheroid component, as we
discussed above.  Because of this, SDSS has been expanded to include a new project, the Sloan Extension for Galactic Understanding and Exploration (SEGUE).  This extension will eventually include 3500 sq. deg. of new
imaging data which is collected in $2.5^\circ$-wide great circles on the sky, but these great circles do not in general
match those laid out on the sky for the SDSS, and they are not adjacent to each other so they do not fill contiguous areas 
of the sky.  The positions of these
stripes were chosen to sparsely sample all directions in the sky that are visible from Apache Point Observatory, and 
include scans at constant Galactic longitude that pass through the Galactic plane.
SEGUE will also obtain $\sim 250,000$ spectra of Galactic stars to study stellar populations and kinematics.

In this paper, we wish to use the photometry derived from imaging of Galactic stars to trace
the three dimensional shape of the Galactic spheroid and the substructure, including tidal streams, contained within
it.  In this demonstration of the algorithm and the results from running it, we will focus on only one well-studied
stripe of SDSS data: stripe 82 on the Celestial Equator.  We use only F turnoff stars in 300 square degrees of this stripe, roughly centered on the
place where the Sgr stream crosses it.  Unlike galaxies,
the distances of stars cannot be inferred from their radial velocities.  By selecting only stars with colors consistent
with spheroid F turnoff stars, we select a stellar population that is not a very good standard candle.  But because
we observe a large number of F turnoff stars we can find their underlying density distribution through statistical
methods.  We will assume the absolute magnitudes of the F turnoff stars in the population have a mean 
of $\bar{M}_{g_0}=4.2$ and a dispersion of $\sigma_{M_{g_0}}=0.6$.  These are reasonable estimates for the 
Sgr dwarf tidal stream \citep{ny06}.

\subsection{The Technique of Maximum Likelihood}

Given a parameterized model and some data generated according to the model
for some ``true'' values of the model parameters, the task of model
estimation is to determine the set of parameters used in generating the
data. Within a Bayesian setting, the model estimation problem
can be reformulated as determining the \emph{a posteriori} most
likely parameters given the data and the model.  In our case we have spatial
positions for a set of stars in the spheroid, and a proposed model for
the spheroid and a tidal stream which passes through it, and we would like to
find the most likely values of the parameters in that model. 

The likelihood of a set of
parameters is the probability of obtaining a particular data set
for a given set of parameters. Via Bayes' theorem, one can decompose  the
\emph{a posteriori} probability of a particular
set of parameters given the data and the model into the product of two terms,
the first being the likelihood of the parameters and the second being the
prior probability of the parameters.
When the prior probability distribution over the parameters is uniform (as is typically assumed), the
\emph{a posteriori} probability is proportional to the likelihood, and so the
maximum likelihood technique may be used to find the most likely model parameters
 [see \citet{fletcher} for more details].

There are two main tasks to implementing the maximum likelihood technique.
The first is to develop the likelihood function,
which measures the probability of observing a particular dataset given a parameterized model.
The second is to maximize the likelihood with respect to the parameters.

In developing the likelihood function, one typically assumes that each data point is
independently generated, hence the likelihood of the dataset is the product
of the likelihoods of each individual data point. Within our setting,
we develop a parameterized model for the background Milky Way and Sagittarius
stellar distributions from which we compute the likelihood.
We prefer to maximize the
logarithm of the likelihood, rather than the likelihood because the computation of
the log-likelihood is numerically more stable.  
We then use the standard technique of optimization via conjugate gradients \citep{fletcher} to
maximize the likelihood and hence determine the most likely model parameters fitting the data.

One advantage of the maximum likelihood framework is that the Hessian
of the log-likelihood (at the maximum) gives the shape of the probability distribution
over the parameters, and hence allows us to determine the \emph{statistical}
error in the estimated model parameters. When the model parameters
are physical quantities of interest, this statistical error translates to
an error bar on the measured (i.e. estimated) physical quantity.

In the remainder of this paper we present a new automated technique that can detect tidal debris in a spatial 
input catalog of stars from one stripe of SDSS data.  Since the first stripe in which the Sgr dwarf tidal 
stream was detected in SDSS data was Stripe 82 \citep{nyetal02}, and it has therefore been the most 
well-studied; this is the section of data on which we test our maximum likelihood.  
Building upon this, we describe a 
technique to extract a catalog of stars that fits the density profile of the debris; this catalog can then be 
used to constrain the dynamical models.  

\section {Algorithm for Fitting Spheroid Substructure\label{algorithm}}

\subsection {Goals of the Algorithm}

Starting with 
a parameterized density model 
for the spheroid, a parameterized density model for the Sgr dwarf tidal stream, the absolute magnitude
distribution for F turnoff stars, and
a dataset of observed $(l,b)$ angular coordinates and apparent magnitudes for color-selected
F turnoff stars in a $2.5^\circ$ stripe through which a tidal stream passes, 
we find the parameters for a tidal stream and stellar spheroid concurrently that give us the
highest likelihood of observing the data.  
This will allow us to tabulate summary statistics for the Sgr dwarf tidal stream, including position,
width, and density.  It will also allow us to test models for the spheroid component of the Milky Way, and ultimately, it will allow us to construct a catalogue of stars that fit the density profile of the Sgr tidal stream by probabilistically separating the stream from the spheroid.

The parameters in our stellar density model describe the position, direction, and width of the tidal stream, 
as well as the shape of the smooth stellar spheroid.  
We start with an initial set of parameters and iterate using a conjugate 
gradient search technique until the optimum values for each parameter given the data set are determined.  From
the parameters which produce the maximum likelihood, we estimate the errors in each parameter.

Once we have found the best parameters, we are able to use a separation algorithm to divide the data into 
two sets of stars, one of which has the density profile of the stream and the other of which has the density profile of the
spheroid.  While we cannot determine which of the individual stars in the data set belong to the stream and which 
belong to the spheroid, we can separate the input catalog into two catalogs that will have the density properties of 
the stream and the spheroid, respectively.  These catalogs will allow for a close comparison of the stream density profile 
with that of simulations of tidal disruption and constrain the dynamical models used in simulations of tidal
disruption.

\S 2.2 gives a brief overview of the algorithm.  \S 2.3 describes the 
construction of the probability density function.  \S 2.4 discusses the
optimizaiton method we use.  \S 2.5 describes the method by which we estimate the errors 
in the parameters.  \S 2.6 discusses the application of a separation 
algorithm to distinguish stream and spheroid stars.  \S 2.7 describes 
the computational time required to run the algorithm and the application 
of parallel processing as a way to improve efficiency.  \S 2.8 discusses 
some limitations and possible enhancements that could be made in the future.  
A thorough discussion of the conjugate gradient and line search optimization 
method can be found in the Appendix.

\subsection{Overview of the Algorithm}

The algorithm starts with initial guesses for the model parameters and an input catalog of observed
F turnoff star positions as $(l,b,g)$, which represent the Galactic longitude, in degrees; Galactic latitude, in degrees; and 
reddening corrected $g$ magnitude for each star.   The apparent magnitudes are corrected for reddening using the
\cite{sfd98} reddening maps, as implemented in SDSS Data Release 6 (DR6).

The essence of the maximum likelihood technique is to create a parameterized probability density function (PDF),
and then find parameters in this function which maximize the likelihood of observing the data.  We first select
a parameterized stellar density function for the spheroid and its substructure.
Since the parameterized models describe the density as a function of space $(X, Y, Z)$, and the observations include only
Galactic coordinates $(l,b)$ and apparent magnitude, $g$, we need to relate spatial position to $(l,b,g)$.
The angular position in the sky is very well known for each star, but it is difficult to determine the distance 
to each star.  We can only estimate the distance from its apparent magnitude, $g$.  If all of the stars in the
sample have the same absolute magnitude, then the conversion of apparent magnitude to distance is trivial.
Unfortunately, this is not a good approximation for F turnoff stars, whose luminosity can deviate from the mean by a factor of $5$ or more, even within the same age and metallicity population.  Instead, we approximate the 
distribution of absolute magnitudes of F turnoff stars as Gaussian.

The likelihood of observing the data is the product of the likelihoods of observing each of the 
stars within the dataset, given the model:
\begin{equation}
\mathcal{L}(\vec{Q}) = \prod_{i=1}^{N} PDF(l_i, b_i, g_i | \vec{Q}),
\end{equation}
where the index $i$ runs over the $N$ stars, $\vec{Q}$ is a vector representing the parameters in the model, and the probability
density function (PDF) is a normalized version of the stellar density function that will be derived in the next section.
Because the individual probabilities are small, we avoid numerical underflow by
maximizing the average logarithm of the likelihood: 
\begin{equation}
\frac{1}{N} \ln{\mathcal{L}(\vec{Q})} = \frac{1}{N} \sum_{i=1}^N \ln{PDF(l_i, b_i, g_i | \vec{Q})},
\end{equation}
which is maximized for the same parameters that the likelihood
is maximized.

We have chosen as input to this first version of the maximum likelihood algorithm a volume of the Galaxy sampled
in a section of stripe 82.  This volume is a piece of a wedge, $2.5^\circ$ wide, with point at the Sun, limited in
near and far end by the apparent magnitude range $16<g_0<22.5$.  For F turnoff stars with absolute magnitude $M_g=4.2$, 
the approximate distance range is $2.3<R<45.7$ kpc from the Sun.  The near and far edges of the volume are fuzzy due to the 
inexact relationship between apparent magnitude and distance, and the variable detection efficiency at the faint end.  We chose this volume of a single stripe as the first application of the algorithm, for we do not have a global model for a general tidal stream; however, if we choose to look at a small enough volume of the stream (as is the intersection of the stream with Stripe 82), we are able to approximate its shape as a cylindrical density with cross-sectional Gaussian fall-off from its axis.  We model the entire tidal stream as a set of these cylindrical stream segments.  This model will not work if the stream is in the plane of the wedge, but we show later that this cylindrical approximation works for this stream in this wedge.
 
The algorithm begins by calculating the average log-likelihood for an initial set of parameters.
A conjugate gradient with line search method is used to find the next set of parameters and the process iterates.  Using this method, each
subsequent set of parameters is guaranteed to produce an equal or higher likelihood in each iteration of the algorithm.
The algorithm continues to generate new sets of parameters until convergence, that is defined by one of a number of conditions: 
either the gradient becomes so small that it would not provide foreseeable improvement, the line search returns negligible movement, or a maximum number of iterations is reached.

\subsection{The Probability Density Function\label{probability}}

\subsubsection{Tidal Stream Model\label{stream}}

In general, tidal streams follow a complex path through the sky.  The stars in the structures may bunch up at apogalacticon and may have a complex cross-sectional density that varies with position along the stream.  However, within a single SDSS stripe through which the stream passes it is reasonable to approximate the path of the stream as linear (see \S3.3).

We model the stream in a piecewise linear fashion with a separate set of parameters for each stripe of data in the SDSS, 
though in principle we could run this algorithm on any $2.5^\circ$-wide great circle, or set of 
$2.5^\circ$-wide great circles on the sky.  The length of the cylinder is limited by the edges of the data 
in one stripe.  The cross section is circularly symmetric with a density that falls off as a Gaussian
with distance from the stream axis.  
Figure 1 depicts the 
data volume and the relationship between the stream parameters and the segment of a cylinder that describes
the stream.  

The SDSS great circle coordinate system $(\mu, \nu)$ is used to measure the angular position on the sky.  
$\nu$ is bounded by $-1.25^{\circ}< \nu < 1.25^{\circ}$ for all stripes, and measures the angular position across the
narrow dimension of the stripe.  
$\nu$ is by definition zero along the center of each stripe of data.  $\mu$ measures the angular distance along 
the great circle swept out by each stripe.  The inclination of each stripe is the maximum angle between that 
stripe and the Celestial Equator.  Thus, $\mu, \nu,$ and the stripe inclination uniquely specify the angular sky coordinates.

The vector $\vec{c}$ points from the Galactic center to the axis of the cylinder.  
Normally, a vector requires three parameters, one for each dimension.  However, since we have the freedom to require
that the position along the cylinder axis to which it points be in the plane that splits the volume in half along
the narrow direction, we reduce that to two parameters.
Enforcing the condition that $\nu = 0$, we are thus able to parameterize the stream center with only a radial 
distance from the Sun, $R$ (in kpc), and the angular position along the stripe, $\mu$ (in degrees).  Therefore, 
$\vec{c}(\mu, R)$ fixes the center point of a piece of tidal debris within an SDSS stripe and lies along the axis of our cylinder.

The unit vector $\hat{a}$ describes the orientation of the axis of the cylinder.  Again, we need only two parameters
because this vector is constrained to be of unit length.  We parameterize this vector using two angles ${\theta}$ 
and ${\phi}$, both in radians.  $\theta$ is the angle between $\hat{a}$ and the Galactic $Z$-axis.
$Z$ is perpendicular to the Galactic plane and points towards the North Galactic Cap.  The azimuthal angle
$\phi$ is measured counter-clockwise around the $Z$ axis, as viewed looking down on the Galaxy from the North Galactic
Pole, starting from the $X$-axis, which points in the direction from the Sun to the Galactic Center.

The last stream parameter, ${\sigma}$ (in kpc), specifies the stream width and is the standard 
deviation of the Gaussian distribution used to describe the density fall off with distance from the cylinder axis.
For a star with spatial coordinates given by $\vec{p}$, the distance, $d$, from the cylinder axis is given by
\begin{equation}
d = \mid(\vec{p} - \vec{c}) - \hat{a} * (\hat{a} \cdot (\vec{p} - \vec{c})\mid.
\end{equation}
In practice, this calculation is performed by first converting each vector to a Galactocentric Cartesian coordinate system.

In summary, we use five parameters to define our cylindrical stream:  ${\mu}$, R, ${\theta}$, ${\phi}$, and ${\sigma}$.  Figure 1 depicts how these parameters are defined with respect to the stripe volume.
Using these parameters, the stellar density of the stream at point $\vec{p}$ is:
\begin{equation}
\rho_{stream}(\vec{p}) \propto e^{-\frac{d^2}{2\sigma^2}}.
\end{equation}
Normalization of the stellar density will be considered once we have assembled the entire probability density function.

\subsubsection{Spheroid}

We model the stellar spheroid with a standard Hernquist profile \citep{h90}:
\begin{eqnarray}
\rho_{spheroid}(\vec{p}) \propto \frac{1}{r (r + r_0)^3}\mbox{, where}\\
r = \sqrt{X^2 + Y^2 + \frac{Z^2}{q^2}}, \nonumber
\end{eqnarray}
and $X$, $Y$, and $Z$ are Cartesian coordinates centered on the Galactic center.  $X$ and $Y$ are in the Galactic plane,
with $X$ directed from the Sun to the Galactic center and $Y$ in the direction of the Solar motion.  $Z$ is perpendicular
to the Galactic plane and points in the direction of the North Galactic Cap.
The spheroid is described by two parameters: $r_0$ which is a core radius, in kpc, and the dimensionless quantity, $q$, 
which is scaling factor in the Z coordinate direction.  For $q<1$ the spheroid
is oblate, for $q=1$ the spheroid is spherically symmetric, and for $q>1$ the spheroid is prolate.

There are many Galactic components (thick disk, bulge, etc) other than the spheroid that we could in principle add to 
our full model.  However, we don't expect many disk stars in our sample because we are fitting data that is too far 
from the plane of the Milky Way, and the color-selected F turnoff stars are bluer than the turnoff of the thick disk 
stellar population.

\subsubsection{Absolute Magnitude Distribution}

In this section we address the fact that stars within our selected color 
range, which are primarily F turnoff stars, do not all have the same 
absolute magnitude.  If we assume that all of the color-selected stars 
have the same absolute magnitude (equal to the mean absolute magnitude 
of the population) when we estimate their distances from the Sun, any 
substructure in the spheroid will appear to be elongated along our line 
of sight.  To account for this, we calculate an ``observed'' spheroid 
spatial density that is elongated along our line of sight by convolution 
of the density model with the absolute magnitude distribution function along our line of sight.
  
We model the distribution of absolute magnitudes as a Gaussian with a 
center of $\bar{M}_{g_0}=4.2$ and dispersion $\sigma_{g_0}=0.6$,  
which is a simplification of the F turnoff star absolute magnitude 
distribution found for globular clusters by \citet{ny06}. Thus, letting 
$M_{g_0}$ be the absolute magnitude, 
\begin{equation}
M_{g_0}=\bar{M}_{g_0}+\Delta M_{g_0},
\end{equation}
where $\bar{M}_{g_0}=4.2$ and $\Delta M_{g_0}$ has a Gaussian distribution
with zero mean and standard deviation $0.6$.
To account for this distribution over
the absolute magnitude, we first derive the probability of observing
a star per unit apparent magnitude per unit solid angle, assuming all 
stars are of absolute magnitude $\bar{M}_{g_0}=4.2$,
and then convolve in apparent magnitude with a Gaussian of dispersion 
$0.6$ centered at zero.  The result is
the probability of observing a star per unit apparent magnitude per 
unit solid angle, with the absolute magnitude
distribution taken into account. 

We will need to refer to several density functions in different
spaces, and so we first lay down some definitions to make this discussion 
clear. We will refer to a density as $\rho_A(x)$ where ${A}$ is the
name of the density which will refer to the coordinate
space in 
which it is defined and ${x}$ refers to a generic variable in this space.
Thus we define the 6 densities:
\begin{equation}
\rho_X(\vec{x}),\quad
\rho_R(R,\Omega),\quad
\rho_{g_{4.2}}(g_{4.2},\Omega),\quad
\rho_{g_0}(g_0,\Omega),\quad
\rho_{\mathcal{R}}(\mathcal{R}(g_0),\Omega),\quad
\rho_{X_c}(\vec{x}),
\end{equation}
where the first three densities are the model in Cartesian, spherical, and apparent magnitude coordinate systems;
\begin{equation}
\rho_X(\vec{x}) = \frac{dV}{dxdydz},\quad
\rho_{R}(R, \Omega) = \frac{dV}{dRd\Omega}\hbox{, and }
\rho_{g_{4.2}}(g_{4.2}, \Omega) = \frac{dV}{dg_{4.2}d\Omega}.
\end{equation}
The densities which take into account the distribution on the absolute
magnitude space are denoted by
${\rho_{g_0}}(g_0, \Omega),{\rho_{\mathcal{R}}(\mathcal{R}(g_0), \Omega)},\hbox{and }\rho_{X_c}(\vec{x})$, where the subscript
${c}$ stands for
convolved.  All coordinate systems are centered at the Sun, the solar position is assumed to be $8.5$ kpc from the Galactic center, and the direction from the Sun to the Galactic center is in the positive $X$ direction.  Here we have made a distinction between $R$ and $\mathcal{R}$:  $R$ denotes the actual distance each star is from the Sun; $\mathcal{R}(g_0)$ denotes the distance a star of apparent magnitude $g_0$ would be from the Sun if it had an absolute magnitude of $\bar{M_{g_0}} = 4.2$.  We have also made the distinction between $g_0$ and $g_{4.2}$:  $g_0$ denotes the actual reddening corrected apparent magnitude of a star.  $g_{4.2}$ denotes the apparent magnitude a star at distance $R$ would be calculated to have if it had an absolute magnitude of $\bar{M}_{g_0} = 4.2$.  We will adopt these definitions for the remainder of this paper.  
 
The Galactocentric Cartesian density, $\rho_X$, is the actual spatial density of stars as described in Equations 4 and 5 for the stream and spheroid, respectively.  We need ${\rho_{X_c}}$, the observed Galactocentric Cartesian density that is elongated along our line of sight to account for Gaussian distribution of absolute magnitudes with dispersion $\Delta M_{g_0}$. We will obtain
$\rho_{X_c}$ through the sequence of transformations
\begin{equation}
\rho_X(\vec{x})\rightarrow
\rho_{R}(R,\Omega)\rightarrow
\rho_{g_{4.2}}(g_{4.2},\Omega)\rightarrow
\rho_{g_0}(g_0,\Omega)\rightarrow
\rho_{\mathcal{R}}(\mathcal{R}(g_0),\Omega)\rightarrow
\rho_{X_c}(\vec{x}).
\end{equation}
The relationship between these densities is determined by the
transformations which take one coordinate space to the other. The 
$X\rightarrow R$ mapping is the well known spherical coordinate transform,
\begin{equation}
\rho_{R}(R,\Omega)=R^2\rho_X(\vec{x}).
\end{equation}

If all of the stars had an absolute magnitude $\bar{M}_{g_0} = 4.2$, then one would measure an apparent magnitude 
\begin{eqnarray}
&g_{4.2} = 4.2 + 5\log_{10}(\frac{R}{10pc}),\hbox{ therefore}&\\
&R = \mathcal{R}(g_{4.2}) = 10^{0.2(g_{4.2}-4.2-10)\hbox{ (kpc), and}}&\nonumber\\
&dR = \frac{\ln 10}{5}Rdg_{4.2}.&\nonumber
\end{eqnarray}
Thus, the relationship
between $\rho_{R}$ and ${\rho_g}$ is given by
\begin{equation}
\rho_{g_{4.2}}(g_{4.2},\Omega)
=\frac{dg_{4.2}}{dR}\rho_R(R,\Omega)
=\frac{\ln 10}{5}R^3\rho_X(\vec{x})
=\frac{\ln 10}{5}\mathcal{R}^3(g_{4.2})\rho_X(\vec{x}).
\end{equation}

The measured $g_0$ is given by 
\begin{equation}
g_0 = g_{4.2} + \Delta M_{g_0}.
\end{equation} 
Since $g_0$ is the sum of independent random variables,
its density
is the convolution of the two densities, i.e., we have that
$\rho_{g_0}(g_0,\Omega)=\rho_g*\rho_{\Delta M_{g_0}}(g_0,\Omega)$,
where the convolution is in the $g$-dimension. Thus,
\begin{equation}
\rho_{g_0}(g_0,\Omega)=
\int_{-\infty}^\infty dg\ \rho_{g_{4.2}}(g,\Omega) \mathcal{N}(g_0-g;u),
\end{equation}
where $\mathcal{N}$ is the Gaussian density function given by:
\begin{equation}
\mathcal{N}(x;u) =  \frac{1}{u \sqrt{2\pi}} e^{\frac{-x^2}{2 u^2}},
\end{equation}
with $u=0.6$.
Switching back from apparent magnitude to spherical coordinates,
\begin{equation}
\rho_{\mathcal{R}}(\mathcal{R}(g_0),\Omega)=\frac{5}{\ln 10}\frac{\rho_{g_0}
(g_0,\Omega)}{\mathcal{R}(g_0)},
\end{equation}
Since the coordinate spaces ${X_c}$ and $\mathcal{R}$ are related by
the spherical coordinate transformation, we may now collect our
results and 
write the convolved
density
$\rho_{X_c}$ in terms of $\rho_X$ as follows:
\begin{eqnarray}
\rho_{X_c}(\vec{x})
&=&\frac{1}{\mathcal{R}^2(g_0)}\rho_{\mathcal{R}}(\mathcal{R}(g_0),\Omega),\\
&=&\frac{5}{\mathcal{R}^3(g_0)\ln 10}\rho_{g_0}(g_0,\Omega),\nonumber\\
&=&\frac{5}{\mathcal{R}^3(g_0)\ln 10}
\int_{-\infty}^\infty dg\ \rho_g(g,\Omega) \cdot
\mathcal{N}(g_0-g;u),\nonumber\\
&=&\frac{1}{\mathcal{R}^3(g_0)}
\int_{-\infty}^\infty dg \mathcal{R}^3(g)\cdot
\rho_X(\vec{x}(\mathcal{R}(g),\Omega)) \cdot
\mathcal{N}(g_0-g;u), \nonumber
\end{eqnarray}
where 
$(\mathcal{R}(g_0),\Omega)$ are the angular coordinates of 
$\vec{x}.$

We have derived the convolved stellar density function
$\rho_{X_c}$, for a generic stellar density $\rho_X$, which could 
represent either the stream or the spheroid densities in our present context.
This convolution is performed separately on the 
stream and spheroid densities, to compute the functions
$\rho^{con}_{stream}(l,b,\mathcal{R}(g_0))$ and 
$\rho^{con}_{spheroid}(l,b,\mathcal{R}(g_0))$.
We us the numerical integration technique of 
Gaussian quadrature \citep{SC} to do the convolution integral.

\subsubsection{Combined Probability Density Function\label{epsilon}}

We are now ready to compute the probability density function for the combination of the stream and spheroid
densities.  To do this we need the stellar densities of the stream and spheroid as derived in \S 2.3.3,
the detection efficiency for finding stars as a function of apparent magnitude, the volume over which the density is defined, and one more parameter, $\epsilon$, which describes the fraction of stars in each of the two components (spheroid and stream).

The detection efficiency function, $\mathcal{E}$, was derived by fitting a sigmoid curve to the detection
efficiency measurement in Figure 2 of \citet{nyetal02}.  The efficiency function accounts for the decrease in 
detection efficiency at faint magnitudes:
\begin{eqnarray}
\mathcal{E}(g_0) = \frac{s_0}{e^{s_1 (g_0 - s_2)} + 1}\hbox{, where}\\
\vec{s} = ( 0.9402, 1.6171, 23.5877 ). \nonumber
\end{eqnarray}

The dimensionless parameter that defines the fraction of the input stars that are in the stream and the fraction that are
in the spheroid is $\epsilon.$  The parameter $\epsilon$ is modeled, as with the stream parameters themselves, separately for each stripe of data that is analyzed.  Thus, the value of $\epsilon$ for a given stripe of data gives only the relative number of stars that comprise the stream as compared to the spheroid for that stripe of data and does not measure the fraction of stars in the stream as a function of position within the Galaxy.

By definition, the fraction of stars in the stream is:
\begin{equation}
P[stream] = \frac{e^{\epsilon}}{1+e^{\epsilon}}.
\end{equation}
Similarly for the spheroid
\begin{equation}
P[spheroid] = \frac{1}{1+e^{\epsilon}}.
\end{equation}
We introduce here the concept of constrained and unconstrained variables.  Instead of $\epsilon$, we could have
defined a variable $f$ which is the fraction in the stream, and then the fraction in the spheroid would be $1-f$.  However,
we would then need to maximize the likelihood subject to the constraint that this parameter must be between
zero and one.  To avoid that constraint, we instead introduced $\epsilon,$ which is not constrained.  If $\epsilon$
is $\infty$, then all of the stars are in the stream, if $\epsilon=0$ the stars are evenly split between the
components, and if $\epsilon=-\infty$ then all of the stars are in the spheroid component. The other parameters are naturally unconstrained and do not need a conversion.

In total, then, we fit eight model parameters: five from the stream fragment in the current stripe 
$(\mu, R, \theta, \phi, \sigma)$, two from the spheroid $(r_0, q)$, and one that specifies the 
fraction of stars in each component $(\epsilon)$.  Together, we will refer to these
parameters as the eight components of the parameter vector $\vec{Q}$.

The probability density function, then, is given by:
\begin{eqnarray}
PDF(l,b,\mathcal{R}(g_0) | \vec{Q}) &=& \frac{e^{\epsilon}}{1 + e^{\epsilon}} 
		   \frac{\mathcal{E}(\mathcal{R}(g_0)) \rho^{con}_{stream}(l,b,\mathcal{R}(g_0) | \vec{Q})}{\int{\mathcal{E}(\mathcal{R}(g_0)) \rho^{con}_{stream}(l,b,\mathcal{R}(g_0) | \vec{Q})}dV} \nonumber\\
& &+ 		   \frac{1}{1 + e^{\epsilon}} 
		   \frac{\mathcal{E}(\mathcal{R}(g_0)) \rho^{con}_{spheroid}(l,b,\mathcal{R}(g_0) | \vec{Q})}{\int{\mathcal{E}(\mathcal{R}(g_0)) \rho^{con}_{spheroid}(l,b,\mathcal{R}(g_0) | \vec{Q})}dV},
\end{eqnarray}
where the integral is over the entire volume probed by the input data. 

Calculation of the volume integrals dominates the runtime of the algorithm.  Since these integrals cannot
be solved analytically for our functions, we
define an integration mesh which divides the total volume into many volume elements.  The edges of the volume
elements are along constant $g$, $\mu$, or $\nu$.  We then calculate 
the spheroid and stream probabilities at the center of each volume element and multiply this by the volume of 
that element.  The values for the spheroid and stream probabilities in all volume elements are then summed and will
be hereafter referred to as the stream and spheroid integrals, respectively.

\subsection{Optimization}
Now that we have defined a probability distribution function and a likelihood measure, we need to find the parameters that maximize that likelihood.  At the beginning of the first iteration of the program, the likelihood is calculated for the input parameters.  Thereafter, a new set of parameters and likelihood is calculated using a conjugate gradient search coupled with a line search technique.  For a detailed description of the conjugate gradient and line search methods employed see the Appendix \S{A.1} and \S{A.2}, respectively.

\subsection{Errors in the Parameter Estimates}

The accuracy of the parameter estimates depends upon the the shape of the likelihood surface at its maximum, which is governed by the number of stars in our input catalog.
The fewer stars, the wider the peak at the maximum and the larger the statistical error.  It is also limited by the accuracy with which we are able
to numerically determine the maximum.  We will address this latter point when we discuss the test data in \S 3.  Here, we
explain how we calculate the accuracy with which we expect to be able to determine each parameter.

We assume that the likelihood surface for the parameters can be reasonably estimated by a Gaussian near the maximum.  
Then, an estimate for the variance of each parameter can be found from the square matrix of second order partial
derivatives of the log-likelihood function, evaluated at the maximum.  This matrix is called a Hessian Matrix, 
$\mathbf{H}$.
The variance matrix $\mathbf{V}$ is the inverse of the Hessian Matrix, normalized by the number of stars in 
the data set:
\begin{equation}
\mathbf{V} = \frac{1}{N} \mathbf{H}^{-1}.
\end{equation}
The square roots of the
diagonal elements of the matrix provide us with the statistical error in the measurement of each parameter.  We compute the Hessian numerically using a finite difference method (as we did the gradient) as follows:
\begin{eqnarray}
H_{ij} &=& \frac{H_{ij}^1 - H_{ij}^2 - H_{ij}^3 + H_{ij}^4}{4h_ih_j}\hbox{, where},\\
H_{ij}^1 &=& \left.\mathcal{L}(Q_j+h_j, Q_i+h_i)\right|_{\hbox{all other $Q_k$ fixed}}, \nonumber\\
H_{ij}^2 &=& \left.\mathcal{L}(Q_j-h_j, Q_i+h_i)\right|_{\hbox{all other $Q_k$ fixed}}, \nonumber\\
H_{ij}^3 &=& \left.\mathcal{L}(Q_j+h_j, Q_i-h_i)\right|_{\hbox{all other $Q_k$ fixed}}, \nonumber\\
H_{ij}^4 &=& \left.\mathcal{L}(Q_j-h_j, Q_i-h_i)\right|_{\hbox{all other $Q_k$ fixed}}.\nonumber
\end{eqnarray}
Here, $Q_j$ is the $j^{th}$ component of parameter vector $\vec{Q}$, similarly for $Q_i$; and $h_j$ is the perturbation value for $Q_j$, similarly for $h_i$.  In practice we use the same $h_k$ to calculate the gradient as the Hessian.  These $h_k$ can be seen in Table 1.  

\subsection{Separation}

Once we have found the values of the parameters in our model, we can use the model to separate the input catalog
of stars into two catalogs, one having the density profile of the stream and the other having the
density profile of the spheroid.  It is possible to determine the probability that a star is in
the stream or in the spheroid, however, it is not possible to determine which stars are in the stream
and which are in the spheroid.  For example, suppose a star selected from the data set is computed to be a stream star with probability 0.6 and that it is a spheroid star with probability 0.4.  We would
put that star in the stream catalog with probability 0.6 and in the spheroid catalog with probability 0.4, but
there is a 48\% chance that the star was put in the wrong file.  The separation has proved very useful for evaluating
the effectiveness of the model fits to the data, and hopefully will be useful for fitting tidal stripping models
to the stream density profiles.  However, it may not be the best method for selecting spectroscopic follow-up targets
to measure the composition and velocities of stream stars, for example.  Spectroscopic follow-up targets might be
better selected based on the probability that it is a stream star, rather than on the catalog to which it was
randomly assigned.

To perform this separation, we calculate for each star the 
probability, $T$, that it was drawn from the stream population given the parameters, where $T$ is defined by 
\begin{equation}
T(l,b,\mathcal{R}(g_0) \mid\vec{Q}) = \frac{S(l,b,\mathcal{R}(g_0)\mid\vec{Q})}
{S(l,b,\mathcal{R}(g_0)\mid\vec{Q}) + B(l,b,\mathcal{R}(g_0)\mid\vec{Q})},
\end{equation}
where $S(l,b,\mathcal{R}(g_0)\mid\vec{Q})$ is the stream portion of Equation 21, and $B(l,b,\mathcal{R}(g_0)\mid\vec{Q})$ is the spheroid portion of Equation 21.

A uniform random number between 0 and 1 is then generated and tested against this probability.  If the probability of 
being drawn from the stream population is greater than that of the random number, then the star is placed into the 
stream star catalog; otherwise, it is placed within the spheroid catalog.  We repeat this process, generating 
a new random number for each star until all stars have been placed into their appropriate catalog.  Thus, we get 
two distinct populations: one that is a smooth spheroid and one that has all the density properties of the stream.

We use this nondeterministic approach of testing the star probability against a random number for extracting the 
stream from the spheroid stars in order to get an accurate representation of the density profiles.  Since the stream 
probability definition is based solely upon the star's distance from the stream axis, 
assigning all stars with a probability greater than a set value to the stream would result in assignment of all 
stars within a certain distance to the stream.  This would not be an accurate representation of the stream as we 
would be carving out a cylinder of stars from the data rather than a set of stars that fit a specific distribution with 
Gaussian cross section in density.

\subsection{Runtime and Distribution}

In determining the likelihood, a trade-off needs to be made in the accuracy of the numerical
integral calculation in Equation 21.  The smaller the step size, 
$\zeta$, in each dimension the more accurate the
integral calculation is, which can lead to faster convergence of the 
search method being used.  However,
increasing the precision of the integral calculation leads to 
polynomially longer calculation time (scaling like
$\zeta^3$).  Thus, the precision to which the integral is calculated is 
the dominant factor in the runtime and
accuracy of our astronomical model.

Testing has shown that in order to generate a meaningful likelihood 
value in stripe 82, a minimum of 1.4 million points within
the stream and spheroid integrals is required.  On a single processor, 
this takes approximately 15 minutes,
assuming there is no magnitude distribution and the non-convolved 
probabilities are used.  As performing a single
conjugate gradient descent can take thousands (or more) integral 
calculations, these long execution times
presented an obstacle to this research.

To generate a meaningful value for the magnitude convolution integral, 
each integral point requires the
calculation of at least 30 surrounding points.  This increases the 
calculation time of a single integral to 7.5
hours.  Again, assuming a bare minimum of 1,000 likelihood evaluations 
for a single conjugate gradient descent
(typically the range is closer to 5,000), evaluating one astronomical 
model would take almost a year on a single
processor.

To alleviate these massive computational requirements, a Generic Maximum Likelihood Evaluator (GMLE\footnote{GMLE is available for download along with more 
information at http://wcl.cs.rpi.edu/gmle}) has
been developed and used \citep{detal07}.  GMLE allows for a likelihood 
calculation to use various distributed
computing environments, such as multiple processors in a cluster, a 
heterogeneous grid of clusters, or even a
supercomputer.

Using this distributed framework with 88 processors on the Rensselaer 
Grid resulted in a 65 times speedup over a
single processor and using 512 nodes of an IBM BlueGene/L system 
resulted in a 148 times speedup. It now takes
only a few days to compute the maximum likelihood parameters for a 
particular model and stripe of data. In both
the grid and supercomputing environments used, overhead of communication 
was very small compared to the
calculation time. Communication time consists of approximately 1-10\% of 
the time to calculate the likelihood and
does not noticeably increase as more processors are used.  With these 
observations, the calculation should scale
to at least 1000 processors on a grid, and 10,000 processors on the IBM 
BlueGene/L.  More precise calculations of
the integral and magnitude convolution will allow the distributed 
calculation to scale to even larger numbers of
processors, due to the increased calculation time assuming relatively 
fixed communication times.

GMLE has also been extended to allow maximum likelihood evaluation over 
the Berkeley Open Infrastructure for Network Computing (BOINC) Internet computing
framework \citep{anderson_boinc_2005}.  This allows users to volunteer 
computing resources by downloading a BOINC client and
attaching to the MilkyWay@Home 
project\footnote{http://milkyway.cs.rpi.edu/milkyway/}.  Currently, over 500
computers from around the world are being volunteered for the project 
and performance is comparable to 512 nodes
of the IBM BlueGene/L.

\subsection{Limitations and Enhancements}

Generically, all enhancements to this algorithm either improve the Galaxy model, increase the quantity of 
data to which the model is fit, or improve the accuracy of speed and accuracy of the convergence.
We plan to improve the algorithm in all of these aspects.

We plan to build up a model of the Milky Way stellar density, starting with the spheroid component.  Over time, we
will be able to include additional streams and additional pieces of the Sagittarius dwarf tidal stream, a more 
sophisticated model of the smooth component of the spheroid, and disk components.  The spheroid substructure is best
fit in a piecewise manner as we have done in this paper, but the smooth components require that more directions be
fit simultaneously.  To be able to fit multiple stripes during a single optimization it must be possible fit multiple debris pieces during a single run as well.  To add additional tidal debris, we need only to add a term for each additional stream to the PDF:
\begin{eqnarray} 
PDF &=& \sum_{i=1}^k 
[\frac{e^{\epsilon_i}}{(1 + \sum_{j=1}^k{e^{\epsilon_j}})}
\frac{\rho^{con}_{stream}(l,b,\mathcal{R}(g_0) | \vec{Q}_{stream_i})}{\int{\rho^{con}_{stream}(l,b,\mathcal{R}(g_0) | \vec{Q}_{stream_i})}dV}] \nonumber\\
& &+\frac{1}{(1 + \sum_{i=1}^k\epsilon_i)} 
\frac{\rho^{con}_{spheroid}(l,b,\mathcal{R}(g_0) | \vec{Q}_{spheroid})}{\int{\rho^{con}_{spheroid}(l,b,\mathcal{R}(g_0) | \vec{Q}_{spheroid})}dV},
\end{eqnarray}
where i and j denote the $i^{th}$ and $j^{th}$ stream of $k$ total streams, respectively.  We add five new parameters, $\vec{Q}_{stream_i}$, and a sixth parameter, $\epsilon_i$, for each new stream segment.  This change to the PDF would allow $k$ pieces of tidal debris to be fit within a single or multiple stripes.  We are considering whether we should enforce continuity conditions between adjacent sections of the tidal stream.  Since it has been suggested that the smooth portion of the spheroid may not be symmetric, we would like also to increase the number of parameters in the smooth spheroid component, following \citet{ny06}.

We are currently fitting only F turnoff stars, and assuming that the absolute magnitude distribution of these stars is
a standard Gaussian for all populations.  Ideally, we would use stars in a wider color range, and develop a probability 
distribution in magnitude and color for each population of stars.  This would increase the number of stars available to
the algorithm and would allow us to use additional information about real structures in the Milky Way - insisting that each population with a 
turnoff also has a main sequence and giant branch, for example.  Implementing this requires significant development not
only in the computation of the likelihood, but also in characterizing Galactic stellar distributions.  Small steps in
improving this aspect of the algorithm are currently underway.

All of the preceding improvements to the algorithm increase the number of fit parameters and the complexity of calculating
the likelihood function.  In parallel with scientific development of the algorithm, we will be improving the speed of 
convergence.  We are experimenting with other methods of finding the maximum, and increasing the number of architectures on
which our software can be run in parallel.  Future architectures that we will use include supercomputers and 
BOINC, which among other products supports the SETI@home application.

\section{Simulated Data \label{test data}}
\subsection{Generation}
To test our algorithm we generated a simulated version of SDSS stripe 82.  The ``True Value" column of Table 2 shows 
the parameter values used for the data generation while Figure 2 shows a plot of the simulated data.  The data set 
contained a total of 205,708 simulated stars; 28,498 (13.85\%)of which are stream stars.  The stream model parameters and star ratio were chosen to be the same as the real data for stripe 82.  The techniques of 
generation are described below.
 
\subsubsection{Stream}
We create the simulated stream using an active generation technique, which means that all points that are generated 
are valid stream stars.  First, we define a set of stream parameters to generate over.  From these 
parameters we then calculate the $\vec{c}$ and $\hat{a}$ vectors defining our stream position and direction.  We 
then generate three random numbers:  a uniform random number that determines where along the stream axis the point 
is, and two Gaussian random numbers (with standard deviation $\sigma$ as defined in the parameters) to define the 
stream cross-sectional coordinates.  The point is then converted to Galactic longitude, latitude and apparent magnitude.  
Thus far, we have assumed a constant absolute magnitude for each star in the stream.  To account for the fact that
there is an absolute magnitude distribution, we add or subtract a small amount from the apparent magnitudes of each
of the stream stars in the dataset.  A Gaussian random variable with standard deviation 0.6 is generated (one for 
each point); this value is added to the apparent magnitude of the generated point.  
Finally, we keep points with a probability given by the efficiency function to produce a data set for a simulated stream.
Stars near the magnitude limit of the data are tossed out of the sample with a probability equal to the efficiency of
our object detection as a function of apparent magnitude in the real data.

\subsubsection{Spheroid}
The simulated spheroid is generated using a rejection sampling technique.  Here we cannot apply the same active generation 
technique we used for the stream because the density distribution in the smooth component of the spheroid is more complex than a
Gaussian.  Instead, we define a rejection technique that allows us to generate over a given stripe.  We cannot 
simply generate values for $\mu$, $\nu$, and $R$, however, because the stars are not uniformly distributed in these variables.  The distribution in $\nu$ varies according to the $\cos(\nu)$, and the distribution in $R$ must take into 
account the growing volume at increasing distance.  $\mu$ may be generated uniformly over the stripe as it 
corresponds to a longitude, and the volume of space in each longitude bin is the same.  
The volume element in a stripe is given by:
\begin{equation}
dV = R^2 \cos(\nu)drd{\mu}d{\nu},
\end{equation}
and the corresponding stripe volume is: 
\begin{equation}
V = \frac{R_{max}^3}{3}(\mu^+ - \mu^-)(\sin(\nu^+) - \sin(\nu^-)),
\end{equation}
where $R_{max}$ is the maximum radial distance, and the positive and negative superscripts refer to the maximum and 
minimum values of that coordinate for the stripe.  These equations are used to define the functions which
allow us to generate in the other two coordinates:  
\begin{eqnarray}
&\nu = \sin^{-1}(u * (\sin(\nu^+) - \sin(\nu^-)) + \sin(\nu^-)),&\\
&R = R_{max} w^{\frac{1}{3}},&\nonumber
\end{eqnarray}
where $u$ and $w$ are uniform random numbers between zero and one.  $\mu$ is generated by a uniform random number between 
the stripe limits.
Once a point is generated, its probability is calculated using the likelihood function developed in \S 2.3.  The 
probability is then divided by the total probability possible for a star.  The total probability possible for a star is simply the maximum value that can be returned by the likelihood function given the current volume and parameter set.  The value of this ratio is then tested against a 
uniform random number between zero and one; the point is put into the data set if the value is greater than the value of the random number.

\subsubsection{Accuracy of Parameter Determination}
Using the method 
described in \S2.5 we calculated the expected accuracy for each parameter.  Since we created the simulated data with 
a known set of parameters, we therefore know the true parameter values of the dataset.  We can 
calculate the Hessian Matrix at the optimum values and subsequently get a set of error bars at the optimum values.  
These error bars therefore give us the accuracy with which we expect to find the values of the parameters.  The values 
of these errors can be found in the column ``Expected Deviation" (the statistical error bar) in Table 2 for the 205,708 star simulated data set.

In addition to the statistical errors, there can be numerical errors driven primarily by the accuracy with which we determine the integrals.  These numerical errors were estimated by determining the accuracy the likelihood can be calculated for a fixed number of points in the numerical integration and then calculating the `error' in each parameter by perturbing a single parameter, holding all others fixed, until a change in the total likelihood is observed that is greater than the minimum accuracy threshold of the numerical integration.  It should be noted that the values quoted as ``Numerical Error" are heuristic errors, meaning they are estimated and not a true error bar.

\subsection{Testing}
We tested our algorithm by letting it optimize to convergence for eight randomized sets 
of input parameters.  Randomized here means a random perturbation of the actual values by 
75 percent of the parameter's value.  Although parameters outside of this range are in 
principle allowed, the likelihood surface for parameter values very far from the correct 
values is so flat, and the gradient so small, that numerical errors dominate our measurement 
of the gradient.  In these eight optimizations:  five optimized to the true values,  
two optimized to a local maxima which has a lower 
likelihood than that of the true values, and one did not converge because it was far
enough from the correct parameter values that the gradient could not be accurately determined.
The local maxima appears to be related to the fact that the
stream crosses the stripe at low inclination, the algorithm sometimes fits 
the data with a much higher inclination angle and larger width to compensate.  All other parameters 
are still optimized to their true values at this local maxima.

The average values of the returned parameters from the five optimizations to the true values 
are presented in the ``Optimized Value" column of Table 2,
and the ``Actual Deviation" is the deference between the ``True Value" and the ``Optimized Value." The returned parameters have very little deviation.  The standard deviation of the five optimizations were calculated to be around an order of magnitude lower than the statistical error calculated from the Hessian.  These can be be seen in the ``Std. Deviations of Optimizations" column of Table 2.
Note that if the theoretical deviation is large for a particular
parameter then near the maximum, the likelihood changes little with the variation of that parameter.  This leads to a larger
``Actual deviation," and can also make it more difficult for the algorithm to numerically find the maximum since the gradient will
be very small.  The errors calculated from the Hessian assume that the maximum of the PDF is found exactly.

As can be seen in Table 2, all of the parameters have ``Actual Deviations" which are smaller than their ``Theoretical Deviations" with the exception of $r_0$.  It has been discovered that the likelihood changes so little compared to perturbations in these parameters when close to the maximum that this parameter is not being calculated accurately enough to reach the ``True Value" seen in Table 2.  The true error bar should be taken as the sum, in quadrature, of the ``Expected Deviation" and ``Numerical Error" columns.  Once this is done the ``Optimized Value" of all parameters are comparable, within the errors, to the ``True Value." 
  
Finally, we ran our separation algorithm upon the average returned parameter values to create two catalogs of stars given those parameters, and we plotted these separately in Figure 3.  Clearly visible is the stream with a Gaussian density fall off in the left plot, while the right plot depicts a smooth Hernquist profile.

We also tested our algorithm on this same simulated dataset of 205,708 stars without using an absolute magnitude correction, thereby simply assuming that all F turnoff stars have an absolute magnitude, $\bar{M}_{g_0} = 4.2$.  This was done to see how important this correction was and to see whether the algorithm could be effectively used without it.  The results showed that the parameters of $\mu$ and $r_0$ were the only values not affected by this correction; the rest of the parameters deviated wildly from their actual values with the stream parameters $R$, $\theta$, and $\phi$ being the worst of the eight parameters reaching upwards of 34$\sigma$ in error.

\subsection{Robustness of Models}
The previous section showed that the algorithm produces the correct answer given that the data 
is drawn from our model.  In this section, we test the algorithm on data that was not generated 
according to the assumed spatial model.  We constructed a reproduction of the Sgr Dwarf Galaxy 
tidal disruption produced by \citet{law05} using a semi-analytic N-body approach.  Using their 
parameters for the Galactic potential and kinematic values, we constructed an orbit for Sagittarius 
using the NEMO Stellar Dynamics Toolbox \citep{teuben}. Specifically, we used the case of a spherical 
dark matter halo, with a velocity dispersion of 114 km/s. We used a Plummer Sphere with one million 
particles as our model for Sagittarius, The sphere was evolved along the orbit 
for 3.18 Gyr until it reached the present position of the Sgr dwarf spheroidal. The resulting plot of 
the disruption of Sagittarius is shown in Figure 4.  It is consistent with the results obtained by \citet{law05}. 

We then selected the volume corresponding to SDSS 
stripe 82 from the N-body simulation and fit it using our model Hernquist profile spheroid and maximum likelihood technique.  
Using the output model parameters found by our 
algorithm, we created a new simulated stream as described in \S{3.1.1}.  Finally, we compared the cross-sections
of this simulated stream with that of a similar stream generated with our density model.

A cross-section of these 
two streams $1$ kpc thick and centered at the returned stream center are plotted, along with a histogram 
of the stars within the cross-section, in Figure 5.  Figure 5 (left panel) depicts the N-body simulation stream 
while Figure 5 (right panel) is that of our simulated stream.  It can be seen that the N-body simulation is 
indeed not Gaussian.  The drawback to testing with an N-body model is that there are no correct model 
parameters with which to compare; but the center, as shown in Figure 5 (left), and the direction, as 
shown in Figure 4, are reasonable.  In order to assess the error in the determination of the stream 
center, one would have to define what is meant by the center of an asymmetric distribution, but the 
algorithm's choice seems reasonable.    

Next, we estimate the curvature of the Sgr tidal stream in stripe 82.  Taking the distance to the center 
of the stream in stripe 82 to be $29$ kpc \citep{netal03} we calculate that the $2.5^{\circ}$-wide 
SDSS stripes would be $1.3$ kpc thick at this distance.  The length of the stream in stripe 82, taking 
an inclination of $30^{\circ}$ from the stripe \citep{fgn05}, is then $2.5$ kpc.  We estimate the radius 
of curvature of the Sgr tidal tail in the orbital plane at this point to be $18$ kpc by fitting a circle 
to the two southern stripes with detections in \citet{netal03}.  The deviation from linear can then be 
found to be $d = 0.06$ kpc at the edge and is very small when compared to the $6$ kpc \citep{fgn05} width 
of the stream or the 6.74 kpc width found in this paper.  The linear approximation is quite reasonable for 
the Sgr stream in stripe 82.

Finally, we tested our algorithm with differing spheroid models.  Because it was easier to change the spheroid
model we fit than to regenerate many sets of spheroid stars, we modified the exponential component of 
the model Hernquist profile the algorithm fits to the data.
Three tests were performed with the exponent of the Hernquist profile set to 3.5, 2.5 and 
2.0 as compared to the value of 3.0 used to generate the data. The results from these tests can be found in
Table 3. In all cases, the stream parameters and the 
epsilon parameter saw little to no change (all optimizations were within the errors calculated above).  As 
expected, the spheroid parameter values were incorrect, but represent the best parameters for the model 
profile fit to the simulated data within our generated volume.

\section{Results From SDSS Stripe 82 \label{stripe 82}}
After validating the algorithm on the simulated data, we then applied the algorithm to the stripe 82 data 
as described in \S2.2.  Figure 6 shows a plot of this data where a piece of the Sgr tidal stream is clearly visible 
as a dense structure on left of the image at $(\alpha, g_0)\approx(31, 21.5)$.  

We selected
from the SDSS SkyServer Data Release 6 all of the sources in stripe 82 that had magnitudes $16<g_0<22.5$, colors of 
$0.1<(g-r)_0<0.3$ and $(u-g)_0>0.4$, were identified as point sources, and were not EDGE or SATURATED.  
Since the selected stars are far from the Galactic plane, the reddening can reasonably be estimated 
from the total amount of dust in that direction in the sky. 
The subscript ``0" indicates that we selected reddening corrected magnitudes from the SDSS databases.  
The cut in (u-g) is made to remove low-redshift QSOs.  We limited the angular length along stripe 82 to 109$^\circ$, 
with right ascension in the range $310^\circ<\alpha<360^\circ$, $0<\alpha<59^\circ$.  Since stripe 82 is centered on 
the Celestial 
Equator, the edges of the stripe are $\pm 1.25^\circ$ in declination and the angle along the stripe is given exactly 
by right ascension.  A small section of data with right ascension $323.2^\circ<\alpha<323.6^\circ$ and 
declination $-1.0^\circ<\delta<-0.7^\circ$ was removed in order to avoid a globular cluster.  The final 
dataset includes 115,907 stars.

As was done for the test data, a number of randomized sets of initial parameters were used (having the same 75\% perturbation) and the average of the parameters of those optimizations that converged to a minima was taken.  In this case ten total optimizations were performed, five optimized to a minima, four to a local minima with lower likelihood than the true minima, and one was again a poor starting point allowing for no optimization.  The Hessian method was then used to calculate the error bars for these parameters at the true minima.  The errors, 
as well as the average parameter values, are tabulated in Table 4.  
There are additional, unknown systematic errors for real data that were not present for the model data.  The quoted
errors assume that the model is an accurate representation of the stellar density function, including our formula
for the smooth spheroid, tidal stream, and absolute magnitude distribution.

The average parameter values were used as input for our separation algorithm to create stream and spheroid catalogs 
for stripe 82.  These catalogs were then plotted in Figure 7.  As can be seen, the Sgr tidal stream has been 
clearly extracted from the spheroid in the left panel while the spheroid remains in the right panel.

We compared the average stream center and direction with
Figure 4 from \citet{nyetal07} in Figure 8 of this paper.  As can be seen in the figure, the stream direction is roughly tangent to the
stream of 2MASS M giants (though at a different distance $-$ there is a difference in scale between distances
measured with the M giants and distances measured with F turnoff stars and RR Lyraes).  Notice also that the stream direction does not lie exactly in the plane of the Sgr dwarf, but is tilted slightly with respect to the $X_{Sgr}-Y_{Sgr}$ plane ($3^{\circ}\pm1^{\circ}$).
We also compare our direction to the directional results of \citet{fgn05},
who estimated the angle between the observational plane and the Sagittarius dwarf stream to be $30^{\circ}$ and 
the angle between the normal to the line of sight towards a point and the tangent of the stream at that point to be 
$10^{\circ}$.  We calculated the former angle from our parameters by taking the angle between the observational plane of stripe 82 and the 
directional vector.  Doing this we get a value of $30^{\circ}\pm3^{\circ}$.  Our value agrees perfectly.  The second angle was calculated by first finding the line of sight vector to
the center of the stream, calculating a normal to that, and then finding the angle between the stream direction and 
that normal.  We found a value of $22.5^{\circ}\pm0.2^{\circ}$.  This difference is explained by the drastically improved accuracy 
of the maximum likelihood algorithm over the estimate by eye for this more difficult angle.  A tabulation of the stream center position and the stream direction in various coordinate systems can be found in Table 5.  Note that the $Cartesian$ $Center$ and $Cartesian$ $Direction$ correspond to the vectors $\vec{c}$ and $\hat{a}$, respectively, that are used in \S2.3.1.

Our results are in reasonable agreement with those of \citet{netal03}, but with much tighter error bars.  The center of the 
stream is shifted from $\alpha=33.99^{\circ}\pm1^{\circ}$ to $\alpha=31.37^{\circ}\pm0.25^{\circ}$.  This shift is 
slightly larger than expected, but not overly so.
Note also that the \citet{nyetal02} estimate of the center in F turnoff stars is $\alpha=33^{\circ}$, which is closer to our value.  Our detection of radial distance, however, corresponds exactly to that
seen in the \citet{netal03}.  Our stream width of $\sigma=2.86$ is equivalent to a FWHM of 6.7 kpc, which is in good agreement with the value of 6 kpc estimated in \citet{fgn05}. 

The stream ratio $\epsilon$ provides a very good separation as shown in Figure 7 and 
corresponds to a detection of approximately 16,000 stream stars out of the total 115,907 F turnoff stars in the
stripe.  By performing a calculation parallelling Equation 10 of \citet{fgn05} we are able to get an estimate of the total stellar mass in stripe 82 as a fraction of the Sgr dwarf itself.  We first calculate the number of stars in Figure 6 of \citet{nyetal02} with $0.1 < (g-r)_0 < 0.3$ and $16.0 < g_0 < 22.5$.  This was calculated to be 2,194 stars.  We then use this in place of the value calculated for F/G type stars in Equation 10 of \citet{fgn05} to get the total number of F turnoff stars currently in the Sgr dwarf itself.  This results in a total of 1,798,700 F turnoff stars in the Sgr dwarf.  Dividing this by the 16,050 stars found to be in the stream in stripe 82 yields that stripe 82 contains 0.9\% of the Sgr dwarf's current F turnoff stars.   

Our spheroid parameters of $q=0.46$ and $r_0=19.4$ are slightly unexpected given the previous work in \citet{belletal07}, which finds that the flattening parameter for the stellar halo closer to $q=0.6$.  However, this does not 
discount that the best likelihood for the stellar halo is obtained from these results, only that they were not the expected.  Further results 
from other data sets and other stripes will be needed to determine the consistency of these results amongst other stripes and confirm such a strong flattening of the stellar halo.

\section{Conclusions}

We have measured the position, direction, distribution, width and number or F turnoff stars of the Sgr tidal 
debris in SDSS stripe 82 as well as the stellar spheroid parameters corresponding to flattening and core 
radius.  Our work has resulted in the improved measurement of the Sgr stream width (FWHM = $6.74\pm0.06$ kpc), 
position [($\alpha$, $\delta$, $R$) = $31.37^{\circ}\pm0.26$, $0.0^{\circ}$, $29.22\pm0.20$ kpc], and 
direction [$(X, Y, Z)_{galactic} = -0.991\pm0.007$ kpc, $-0.042\pm0.033$ kpc, $0.127\pm0.046$ kpc].  
We measured the number of F turnoff stars in the stream in SDSS stripe 82 ($13.86\pm0.06$\% of 115,907).  
There are $16,050$ Sgr F turnoff stars in stripe 82 compared to $1.8\times10^6$ turnoff stars in the Sgr 
dwarf itself ($0.9\%$).  We also calculated values for the stellar spheroid parameters:  the flattening 
parameter ($q = 0.46\pm0.024$) and the core radius ($r_0 = 19.40\pm0.59$).  While these spheroid parameters 
are not the expected values, they are intriguing and deserve further study.  Finally, we were able to generate 
separate catalogs of stream stars and spheroid stars for the F turnoff stars in stripe 82.  These catalogs, 
while not a true representation of the actual stars in the stream or spheroid, provide the correct density 
characteristics of these structures and can be compared with the density profiles of N-body simulations.

The results were calculated using a new method to study tidal debris that is more robust than previous methods.  This method will prove valuable to the study of tidal disruption and Galactic structure as it provides a means to accurately and efficiently study many variables used to model the spheroid and tidal debris.  Also, we are able to extract a catalog of stream stars which can be used to better constrain models of the Sgr dwarf disruption by providing actual stream stellar positions to compare with the data.   

The algorithm itself is quite flexible.  The stream and spheroid models can be easily changed should more accurate ones be discovered.  The new model definitions would simply replace the existing probability code with no other changes.  Also, with minor changes to the volume definitions it would be possible to use data from surveys other than the SDSS as an input catalog.  With some revisions, the code could be adapted to search for multiple debris and use multiple volumes to provide more accurate results.  

\acknowledgments
This publication is based upon work supported by the National Science Foundation under Grant No. SEI(AST)-0612213.  Additional
support was provided by the NSF grants AST 0607618 and CNS 0323324, NASA New York Space Grant, and John A. Huberty.  We thank Brian Yanny for his help in the extraction of data and overall troubleshooting he did to aid us in this work.

We use data from the Sloan Digital Sky Survey.
Funding for the SDSS and SDSS-II has been provided by the Alfred P. Sloan Foundation, the Participating 
Institutions, the National Science Foundation, the U.S. Department of Energy, the National Aeronautics and 
Space Administration, the Japanese Monbukagakusho, the Max Planck Society, and the Higher Education Funding 
Council for England. The SDSS Web Site is http://www.sdss.org/.

The SDSS is managed by the Astrophysical Research Consortium for the Participating Institutions. The Participating 
Institutions are the American Museum of Natural History, Astrophysical Institute Potsdam, University of Basel, 
University of Cambridge, Case Western Reserve University, University of Chicago, Drexel University, Fermilab, 
the Institute for Advanced Study, the Japan Participation Group, Johns Hopkins University, the Joint Institute 
for Nuclear Astrophysics, the Kavli Institute for Particle Astrophysics and Cosmology, the Korean Scientist 
Group, the Chinese Academy of Sciences (LAMOST), Los Alamos National Laboratory, the Max-Planck-Institute 
for Astronomy (MPIA), the Max-Planck-Institute for Astrophysics (MPA), New Mexico State University, Ohio 
State University, University of Pittsburgh, University of Portsmouth, Princeton University, the United States 
Naval Observatory, and the University of Washington.

\clearpage
\appendix
\section*{APPENDIX}
\section{Optimization Technique} 
\subsection{Conjugate Gradient}
The conjugate gradient method is similar to the gradient method in that we perturb the current parameter value, $Q_i$, by a small amount, $h_i$, and get a direction for the change in the parameter space.  However, unlike the gradient method, the conjugate gradient method enforces the condition that each direction is conjugate to the previous direction.  This means that as we move along the new search direction, the component of the gradient parallel to the previous search direction must remain zero.  This condition speeds up the optimization by using search directions that are non-interfering.  A further discussion may be found in \citet{fletcher}.

The $i^{th}$ component of the gradient vector, $\vec{G}$, is calculated numerically by perturbing $Q_i$, holding all other $Q_j$ fixed:
\begin{equation}
\left.G_i = \frac{\mathcal{L}(Q_i + h_i) - \mathcal{L}(Q_i - h_i)}{2 h_i}\right|_{\hbox{all other $Q_j$ fixed}},
\end{equation}
where $G_i$ is the $i^{th}$ component of the gradient vector, $Q_i$ is the $i^{th}$ parameter, $h_i$ is the perturbation amount for that parameter, and $\mathcal{L}$ is the likelihood function from \S 2.2.

This gradient is used to calculate the direction, $\vec{D}$, for the first iteration:  
\begin{equation}
\vec{D}_1 = \vec{G}_1.
\end{equation}
We take the positive value of the gradient, for we want to maximize the likelihood. Should we want to perform a simple gradient ascent we would do this same calculation for every iteration and use solely the direction calculated above; however, we wish to apply the conjugate gradient technique which always calculates a direction conjugate to the previous directions.  This is done by calculating a multiplier, $B_i$, based upon the current gradient and the previous gradient, as;  
\begin{equation}
B_i = \frac{\vec{G}_{i} \cdot (\vec{G}_{i} - \vec{G}_{i-1})}{\vec{G}_{i-1}^2},
\end{equation}
where $\vec{G}_i$ denotes the gradient vector for the $i^{th}$ iteration.  This value is then used to calculate the new conjugate gradient direction:
\begin{equation}
\vec{D}_{i} = \vec{G}_{i} + B_i * \vec{D}_{i-1},
\end{equation}
where $\vec{D}_{i}$ denotes the direction vector of the $i^{th}$ iteration (the direction to move the current parameters) and $\vec{G}_{i}$ is the gradient of the current parameters, $Q_i$, at the the $i^{th}$ iteration.    

Due to the large variety of parameters that we utilize, all on different scales, we do not use one value for our perturbation value, $h$, for it could potentially produce poor estimates for the gradient if the perturbation is too large or small for
a given parameter.  If $h$ is too large, then the computed gradient is not an accurate approximation of the true gradient at the current point; if $h$ is too small then we may encounter computational errors.  We therefore choose an appropriate value for each parameter independently.  Table 1 shows the perturbation values for each parameter.  These numbers were chosen in two ways.  The values of the gradient were calculated at a number of differing perturbation values in order to determine how much the gradient changed.  Also, plots of all parameter spaces were generated in order to determine the distribution of each parameter.  Combining these two techniques we were able to determine a value for the perturbation for each parameter that would be small enough to provide a very accurate gradient calculation, but also large enough to overcome any anomalous behavior within the parameter's distribution.

\subsection{Line Search}
The line search technique seeks to determine the value, $\alpha$, that minimizes the function
\begin{equation}
\Phi(\alpha) = \mathcal{L}(\vec{Q}_k + \alpha \vec{D}_k),
\end{equation}
where $\vec{D}_k$ is the direction and $\vec{Q}_k$ is the current set of parameters.  After finding $\alpha^*$ which minimizes $\Phi(\alpha)$, we then update
\begin{equation}
\vec{Q}_{k+1} \longleftarrow \vec{Q}_k + \alpha^* \vec{D}_k.
\end{equation}  

A bracketing method must be employed first to ensure that the minimum of the function is within the range we are 
searching.  This is done by calculating the likelihood at three points along the current search direction:  the current parameter values plus zero, one, and 
two times $\vec{D}_k$.  This corresponds to the current position, the current position plus the full direction, and the current position plus twice the full direction.  If the middle 
point does not have a likelihood greater than the two endpoints, the endpoint with the highest probability becomes 
the new center point, the center point the end, and a new third point is calculated by expanding to two times the 
current factor times direction and the new likelihood is calculated.  Iteration continues until the midpoint has 
a greater likelihood than both endpoints.

These three points are then passed to the line search algorithm which will use them to find the peak of the parabolic function defined by moving along the current directional vector and calculating the likelihood at points along this path.  The line search iterates as follows:  first a guess is made for the value of $\alpha$ by fitting a parabola via the calculation:
\begin{eqnarray}
u &=& a_1^2*(L_2-L_3) + a_2^2*(L_3-L_1) + a_3^2*(L_1-L_2),\\
b &=& a_1*(L_2-L_3) + a_2*(L_3-L_1) + a_3*(L_1-L_2),\nonumber\\
\alpha &=& 0.5*\frac{u}{b},\nonumber
\end{eqnarray}
where $a_1$, $a_2$, and $a_3$ are the factors multiplied by the current direction that produce the parameters with likelihood $L_1$, $L_2$, and $L_3$, respectively.  For the first iteration, $a_1$, $a_2$, and $a_3$ are the values returned from the bracketing method.  $\alpha$ is then the current guess for the value to minimize Equation A5.  The likelihood $L_{\alpha}$ is then calculated using the parameters generated using $\alpha$.  The new order of the three points is: 
\begin{eqnarray}
\mbox{if}: \alpha > a_2 \mbox{ and }  L_{\alpha} > L_2; \mbox{then}: 1,2,\alpha,\\   
\mbox{if}: \alpha > a_2 \mbox{ and }  L_{\alpha} < L_2; \mbox{then}: 2,\alpha,1,\nonumber\\
\mbox{if}: \alpha < a_2 \mbox{ and }  L_{\alpha} > L_2; \mbox{then}: \alpha,2,3,\nonumber\\
\mbox{if}: \alpha < a_2 \mbox{ and }  L_{\alpha} < L_2; \mbox{then}: 1,\alpha,2.\nonumber
\end{eqnarray}
In short, the distance along the search direction is reduced based upon the value of $\alpha$ and its corresponding likelihood.  These new points are then used to calculate a new guess for $\alpha$ using Equations A6 through A8.  Iteration continues for a set number of iterations until returning the optimum value of $\alpha$; we currently use three.  

Once we have calculated the change in the parameters, utilizing the conjugate gradient and line search methods, 
we then update the value of the probability for the fit using the same method as discussed in \S 2.3 and 
continue to maximize our probability through the use of the conjugate gradient and line search methods.
The algorithm stops iterating and returns the current values when the 
value of the largest component of the conjugate gradient drop below a set threshold, the line search returns a value that would cause negligible movement, or the algorithm has completed a maximum number of iterations.

\clearpage
\figcaption {Stripe and stream parameter definitions.  The segment of a stream which passes through an SDSS stripe
is cylindrical within an individual stripe, with density that falls off as a Gaussian with distance from its axis.  The coordinates $\mu$, $\nu$, and $r$ are used to 
define SDSS stripes.  We adopt these coordinates to define a vector, $\vec{c}(\mu, R)$ ($\nu=0$), which points to 
the center of the stream from the Galactic center.  We then define a stream directional vector $\hat{a}(\theta, \phi)$ 
of unit length.  Finally, we define the stream width as $\sigma$, which is the standard deviation of the Gaussian 
that defines the density fall-off of the stream.
\label{streamDef}} 

\figcaption {Simulated data wedge plot.  We show a Sun-centered density plot of the 205,708 stars generated to mimic SDSS 
stripe 82.  $g_0$ is labeled along the radial spokes; the circle indicates where $g_0 = 23$.  The angle $\mu$ in degrees is marked along this circle.  For stripe 82 $\mu = \alpha$ and $\nu = \delta$.  The simulated stream is easily discerible at ($\alpha$, $g_0$) = (33.4, 21.4).
\label{testdata}}

\figcaption{Simulated stream and spheroid stars after separation.  Here we 
have plotted the same 205,708 star simulated data set similarly to Figure 2.  
However, we have separately plotted the stream (left) and spheroid (right) catalogs returned 
from running our separation algorithm.  The spheroid is recovered as a smooth Hernquist function after the removal 
of the stream component.\label{testdataSeparation}}

\figcaption{Plot of the disruption of the one million particle simulated 
Sagittarius Dwarf with a spherical dark matter halo (q = 1), 
and a velocity dispersion of $v_{halo} = 114$ km/s in the Sgr XY orbital 
plane. To reduce crowding, the stream was sampled one in one hundred.  
The thin line is the future orbit of the core, while the dotted line is the past orbit. The past orbit is not closed, but in fact follows the lower trailing stream.  The arrow depicts the center and direction returned by our algorithm optimization.\label{sgrNBodySim}}

\figcaption{Cross-sections of Sgr N-body simulation stream (left panel) 
and simulated stream (right panel).  The simulated stream matches our stream model and was generated from
the parameters fit to the N-body simulation.  
The upper figure in both panels show the 1 kpc thick cross-section of the respective data set.  
The cross-sections are centered at the best-fit value of the optimization of the center in the N-body simulation 
from Figure 4.  Here the axes are perpendicular to the stream direction.  The X-axis is  
$0.053X + 0.055Y + 0.997Z$ and the Y-axis is $0.055X + 0.997Y - 0.058Z$, where X,Y,Z are Galactocentric 
Cartesian coordinates with the Sun at X = -8.5 kpc and moving in the direction of positive Y.
The lower figure of both panels is a histogram of those stars within the cross-section binned along the  
X-axis.  The heavy dashed line shows a Gaussian 
distribution with standard deviation given by $\sigma$ from the fit to the N-body simulation.  
Note that the cross-section of the N-body simulation is non-Gaussian and somewhat asymmetric.  
Also note the model simulated stream is well fit by a Gaussian.  The density distributions in the left and 
right panels are not the same, however, the algorithm still fits a reasonable center for the non-Gaussian 
N-body data set.\label{crossSection}}

\figcaption{F turnoff stars within SDSS stripe 82.  Here we plot, as in the Figure 2, F turnoff stars with 
color cut 0.1 $< (g-r)_0 >$ 0.3 and $(u-g)_0 > 0.4$ that are within the volume defined by 
$310 < \alpha < 419$, $\pm1.25=\delta$, and $16 < g_0 < 22.5$ that were detected in SDSS stripe 82.  Sgr tidal debris 
is visible around ($\alpha, g_0$) = (31, 21.5). \label{sdss82}}

\figcaption{F turnoff stars within SDSS stripe 82 after separation.  Here we have plotted the data 
from Figure 6 after being separated into two distinct catalogs: one for stream and one for spheroid.  The 
stream (left) has clearly been extracted from the spheroid (right).\label{82seperation}}

\figcaption{Plot of the stream center and direction in the Sgr dwarf plane compared to other Sgr detections.
Here we reproduce Figure 4 of \citet{nyetal07} which displays detections of the Sgr stream in A-colored stars in eleven stripes, where the filled circles and larger squares represent leading debris, the open circles trailing debris, and the smaller squares Sgr debris on the opposing side of the Sgr orbital plane.  These are plotted along with the positions of 2MASS M giants from Figure 11 of \citet{mswo03}.  The arrow shows our improved measurement of the position and direction in one place on the Sgr tidal stream.  The length of the directional vector is arbitrary representing only the projection of an elongated unit vector onto the respective planes.  Note that the direction is tangent to the 2MASS M star data, and plausibly on a smooth path from stripe 86 (the open circle just to the right of our detection for stripe 82) to stripe 27 (the open circle farthest to the left in the lower figure).  
\label{virgoFig}} 

\clearpage
\begin{deluxetable}{rrrrrrrr}
\tabletypesize{\scriptsize}
\tablecolumns{5}
\footnotesize
\tablecaption{Perturbation values used for the gradient and Hessian}
\tablewidth{0pt}
\tablehead{
\colhead{Parameter} & \colhead{h}}

\startdata
$\mu$ (deg) & $3 \cdot 10^{-5}$\\
$R$ (kpc) & $4 \cdot 10^{-5}$ \\
$\theta$ (rad) & $6 \cdot 10^{-5}$\\
$\phi$ (rad) & $4 \cdot 10^{-5}$\\
$\sigma$ (kpc) & $4 \cdot 10^{-6}$\\
$\epsilon$ & $1 \cdot 10^{-6}$\\
$q$ &  $4 \cdot 10^{-6}$\\
$r_0$ (kpc) & $8 \cdot 10^{-4}$\\
\enddata
\end{deluxetable}

\clearpage
\begin{deluxetable}{rrrrrrrr}
\tabletypesize{\scriptsize}
\tablecolumns{7}
\footnotesize
\tablecaption{Results for the simulated data set of 205,708 stars}
\tablewidth{0pt}
\tablehead{
\colhead{Parameter} & \colhead{True} & \colhead{Expected} & \colhead{Numerical} & \colhead{Optimized} & \colhead{Actual} & \colhead{Std. Deviation}\\
\colhead{} & \colhead{Value} & \colhead{Deviation} & \colhead{Error} & \colhead{Value} & \colhead{Deviation} & \colhead{of Optimizations}}

\startdata
$\mu$ (deg) & 31.361 & 0.233 & 0.05 & 31.443 & 0.082 & 0.064\\
$R$ (kpc) & 29.228 & 0.167 & 0.04 & 29.217 & 0.011 & 0.010\\
$\theta$ (rad) & 1.445 & 0.032 & 0.003 & 1.421 & 0.024 & 0.0005\\
$\phi$ (rad) & 3.186 & 0.049 & 0.001 & 3.182 & 0.004 & 0.002\\
$\sigma$ (kpc) & 2.854 & 0.033 & 0.015 & 2.858 & 0.004 & 0.009\\
$\epsilon$ & -1.828\tablenotemark{1} & 0.005 & 0.002 & -1.833 & 0.005 & 0.005\\
$q$ &  0.670 & 0.013 & 0.000 & 0.671 & 0.001 & 0.0004\\
$r_0$ (kpc) & 13.500 & 0.276 & 0.15 & 13.917 & 0.417 & 0.016 \\
\enddata
\tablenotetext{1}{Corresponds to a stream star ratio of 13.85 percent: 28,498 in stream; 177,210 in spheroid.}
\end{deluxetable}

\clearpage
\begin{deluxetable}{rrrrrrrr}
\tabletypesize{\scriptsize}
\tablecolumns{7}
\footnotesize
\tablecaption{Stream fitting results for tests with incorrect spheroid model.}
\tablewidth{0pt}
\tablehead{
\colhead{Parameter} & \colhead{True} & \colhead{Expected} & \colhead{$\alpha = 3.5$} & \colhead{$\alpha = 2.5$} & \colhead{$\alpha = 2.0$}\\
\colhead{} & \colhead{Value} & \colhead{Deviation} & \colhead{Optimization} & \colhead{Optimization} & \colhead{Optimization}}

\startdata
$\mu$ (deg) & 31.361 & 0.233 & 31.449 & 31.526 & 31.457\\
$R$ (kpc) & 29.228 & 0.167 & 29.094 & 29.326 & 29.108\\
$\theta$ (rad) & 1.445 & 0.032 & 1.426 & 1.452 & 1.437\\
$\phi$ (rad) & 3.186 & 0.049 & 3.172 & 3.168 & 3.160\\
$\sigma$ (kpc) & 2.854 & 0.033 & 2.869 & 2.848 & 2.865\\
\enddata
\end{deluxetable}

\clearpage
\begin{deluxetable}{rrrrrrrr}
\tabletypesize{\scriptsize}
\tablecolumns{4}
\footnotesize
\tablecaption{Results for the 115,907 F turnoff stars in SDSS Stripe 82}
\tablewidth{0pt}
\tablehead{
\colhead{Parameter} & \colhead{Optimized} & \colhead{Statistical} & \colhead{Numerical} & \colhead{Std. Deviation}\\
\colhead{} & \colhead{Value} & \colhead{Error} & \colhead{Error} & \colhead{of Optimizations}}

\startdata
$\mu$ (deg) & 31.373  & 0.244 & 0.08 & 0.008\\
$R$ (kpc) & 29.218 & 0.184 & 0.07 & 0.012\\
$\theta$ (rad) & 1.444 & 0.044 & 0.01 & 0.001\\
$\phi$ (rad) & 3.184 & 0.034 & 0.008 & 0.002\\
$\sigma$ (kpc) & 2.862 & 0.025 & 0.008 & 0.009\\
$\epsilon$ & -1.827\tablenotemark{1} & 0.005 & 0.001 & 0.000\\
$q$ & 0.458 & 0.023 & 0.005 & 0.001\\
$r_0$ (kpc) & 19.404 & 0.581 & 0.09 & 0.051\\
\enddata
\tablenotetext{1}{Corresponds to a stream star ratio of 13.86 percent of the 115,907 F turnoff stars; about 16,000 stream stars.}
\end{deluxetable}

\clearpage
\begin{deluxetable}{rrrrrrrr}
\tabletypesize{\scriptsize}
\tablecolumns{6}
\footnotesize
\tablecaption{Table of the stream center and direction as calculated for different coordinate systems.  All distances are in kpc and angles are in degrees.}
\tablewidth{0pt}
\tablehead{
\colhead{Physical} & \colhead{Coordinate} & \colhead{Value 1} & \colhead{Value 2} & \colhead{Value 3}\\
\colhead{Quantity} & \colhead{system} & \colhead{} &\colhead{} & \colhead{}}

\startdata
$Center$ & $Equatorial$ ($\alpha$, $\delta$, $R$) & 31.37 & 0.0 & 29.22\tablenotemark{1}\\
$Center$ & $Equatorial$ ($\alpha$, $\delta$, $g_0$) & 31.36 & 0.0 & 21.53\\
$Center$ & $Galactic$ ($l$, $b$, $R$) & 159.223 & -57.558 & 29.22\tablenotemark{1}\\
$Center$ & $Galactic$ ($l$, $b$, $g_0$) & 159.223 & -57.558 & 21.53\\
$Center$ & $Cartesian$ ($X$, $Y$, $Z$) & -23.154 & 5.560 & -24.658\\
$Direction$ & $Cartesian$ ($X$, $Y$, $Z$) & -0.991 & -0.042 & 0.127\\
\enddata
\tablenotetext{1}{Assuming an absolute magnitude of $\bar{M}_{g_0} = 4.2$.}
\end{deluxetable}

\clearpage
\setcounter{page}{1}

\plotone{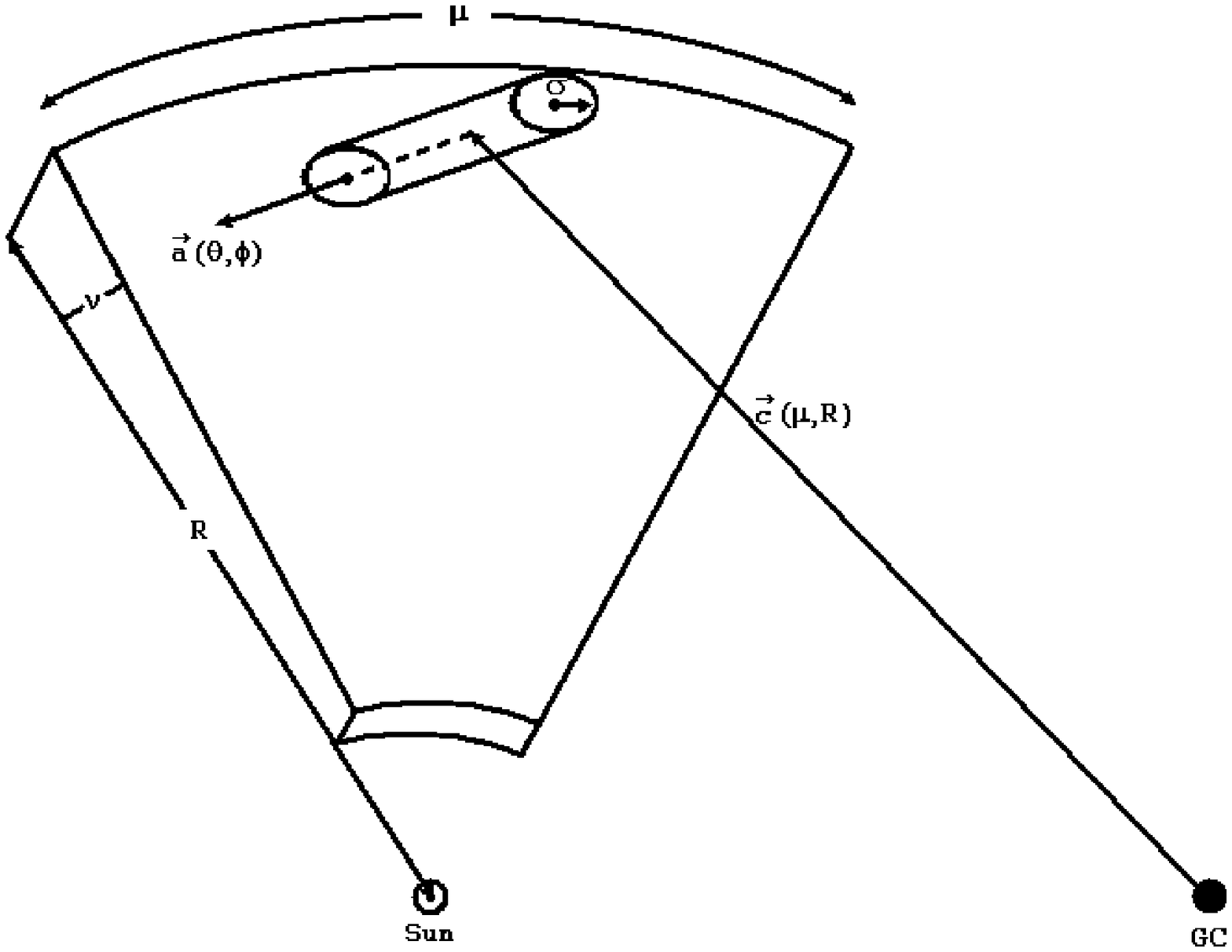}

\plotone{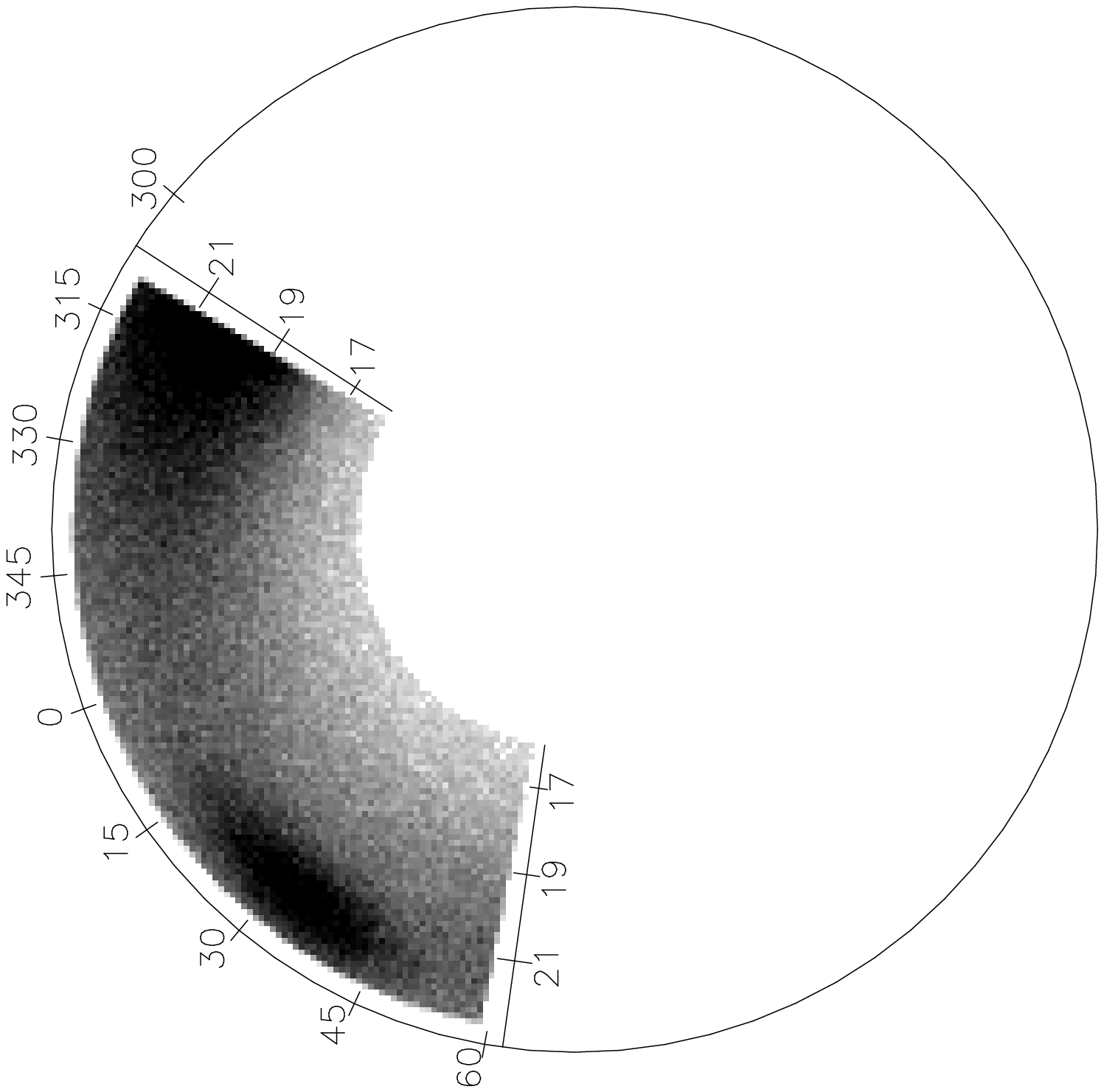}

\plotone{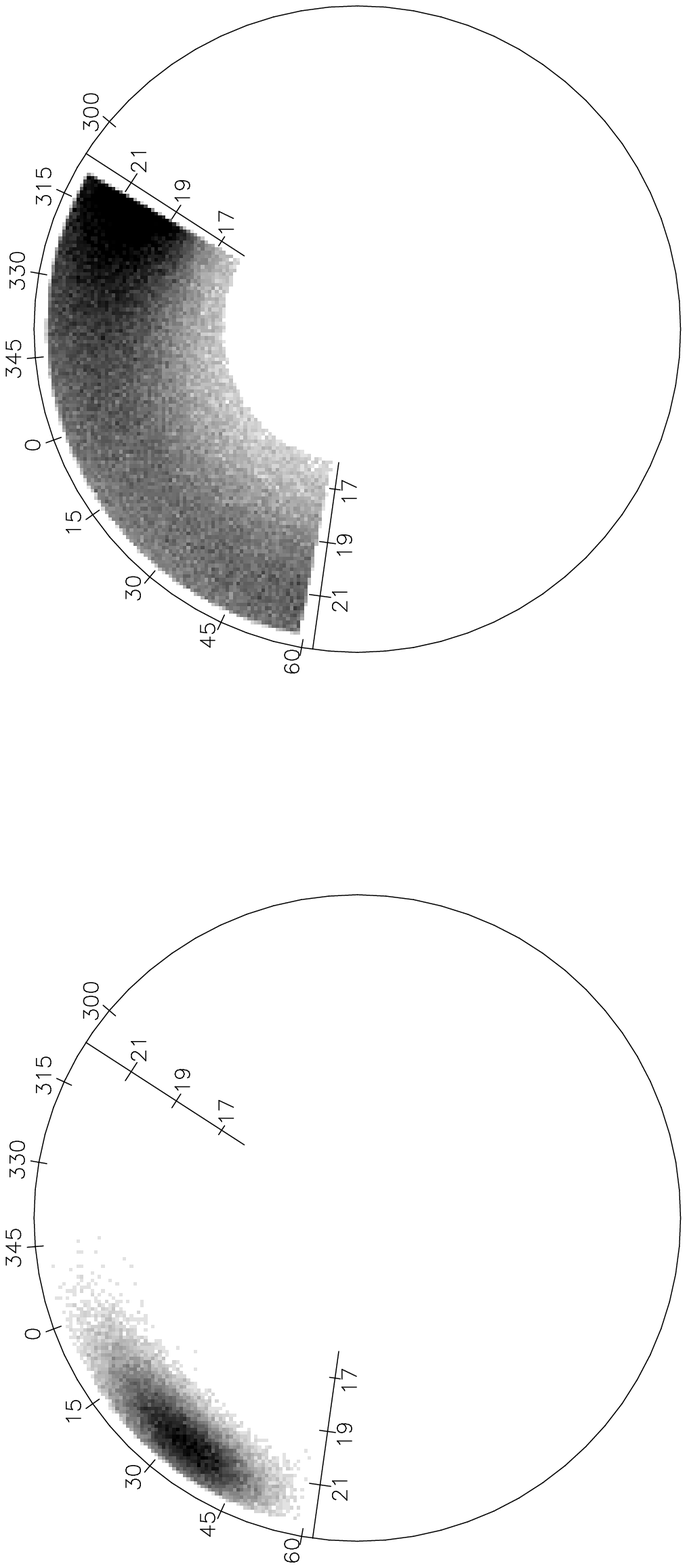}

\plotone{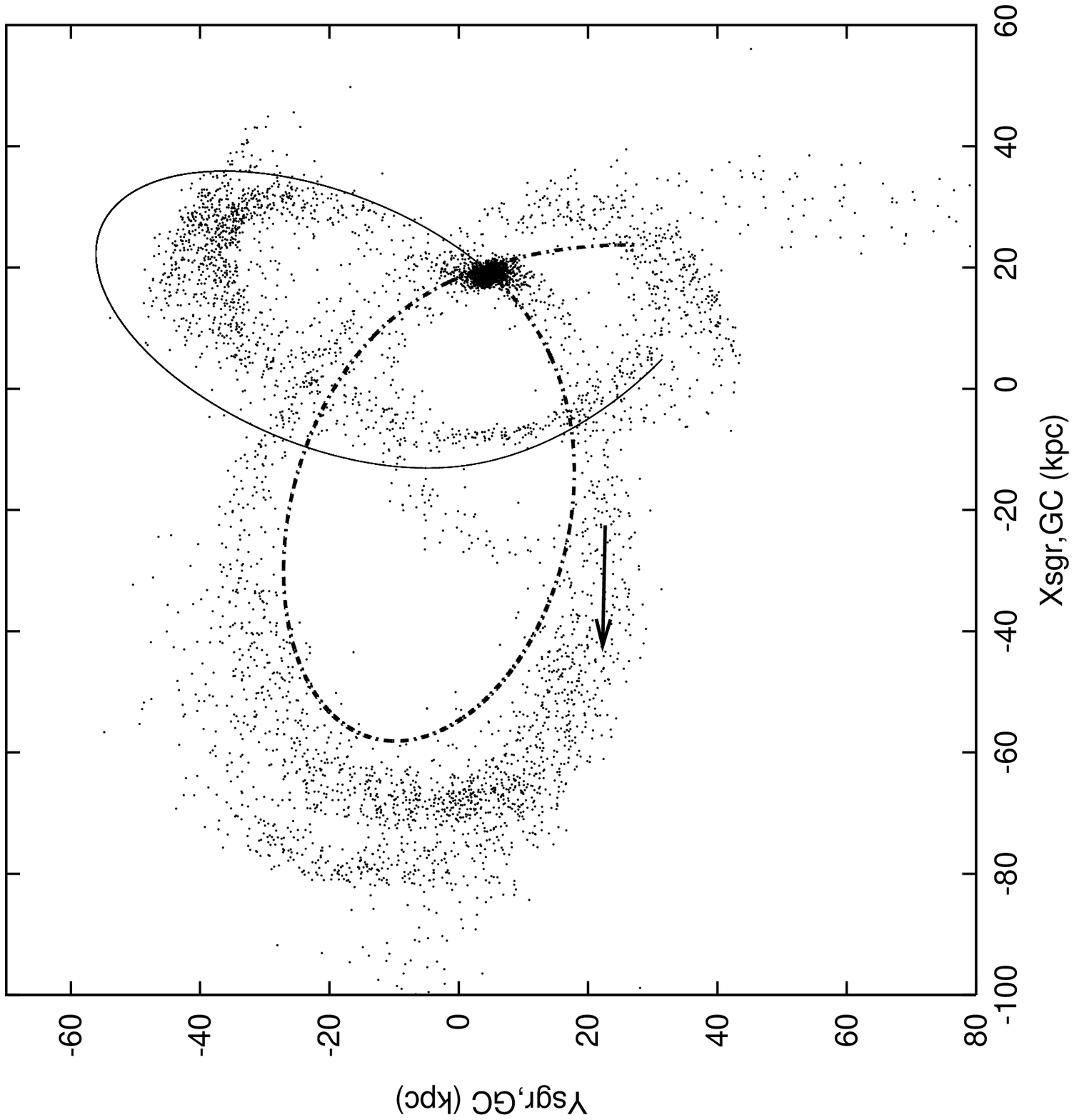}

\plotone{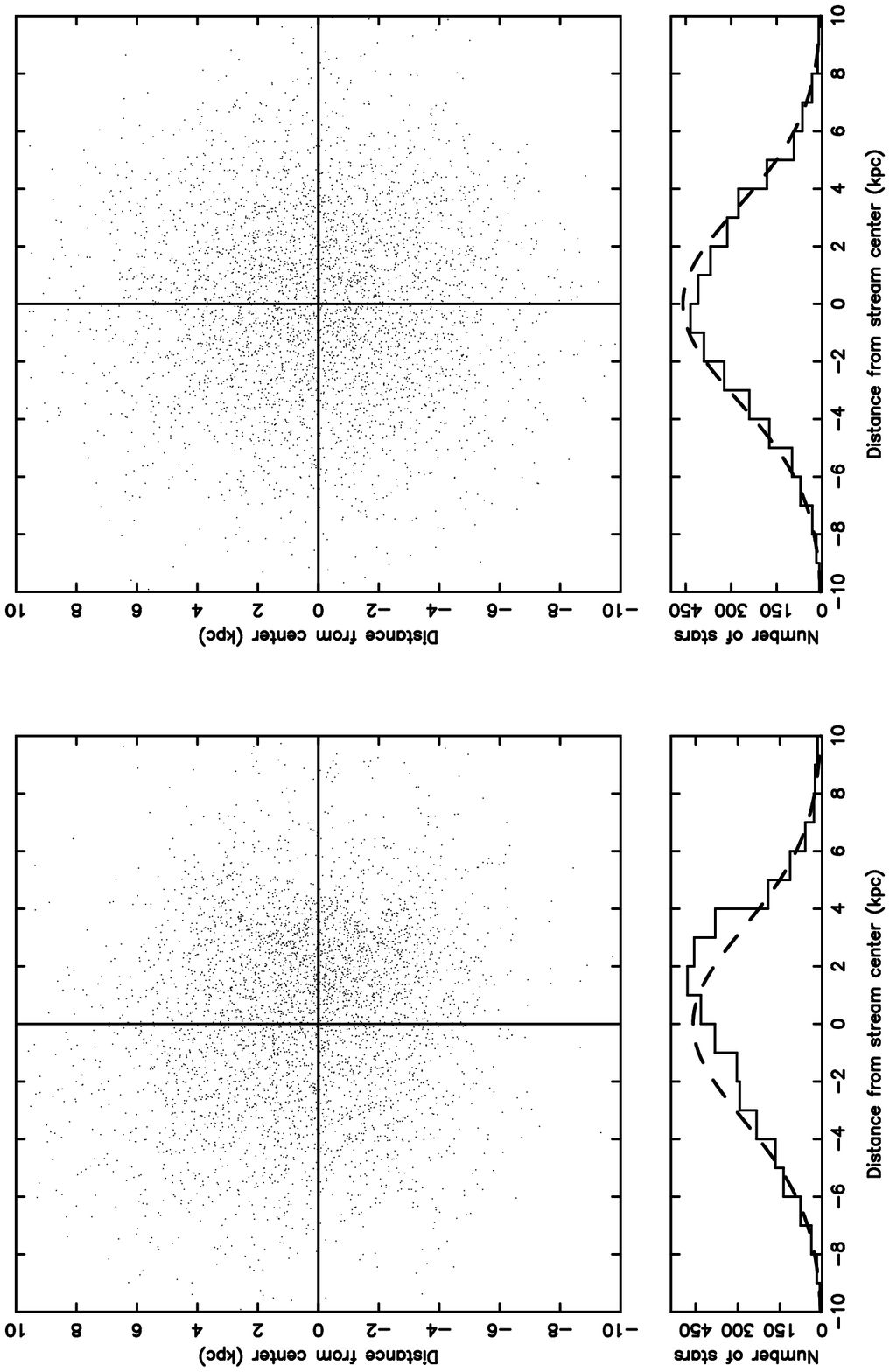}

\plotone{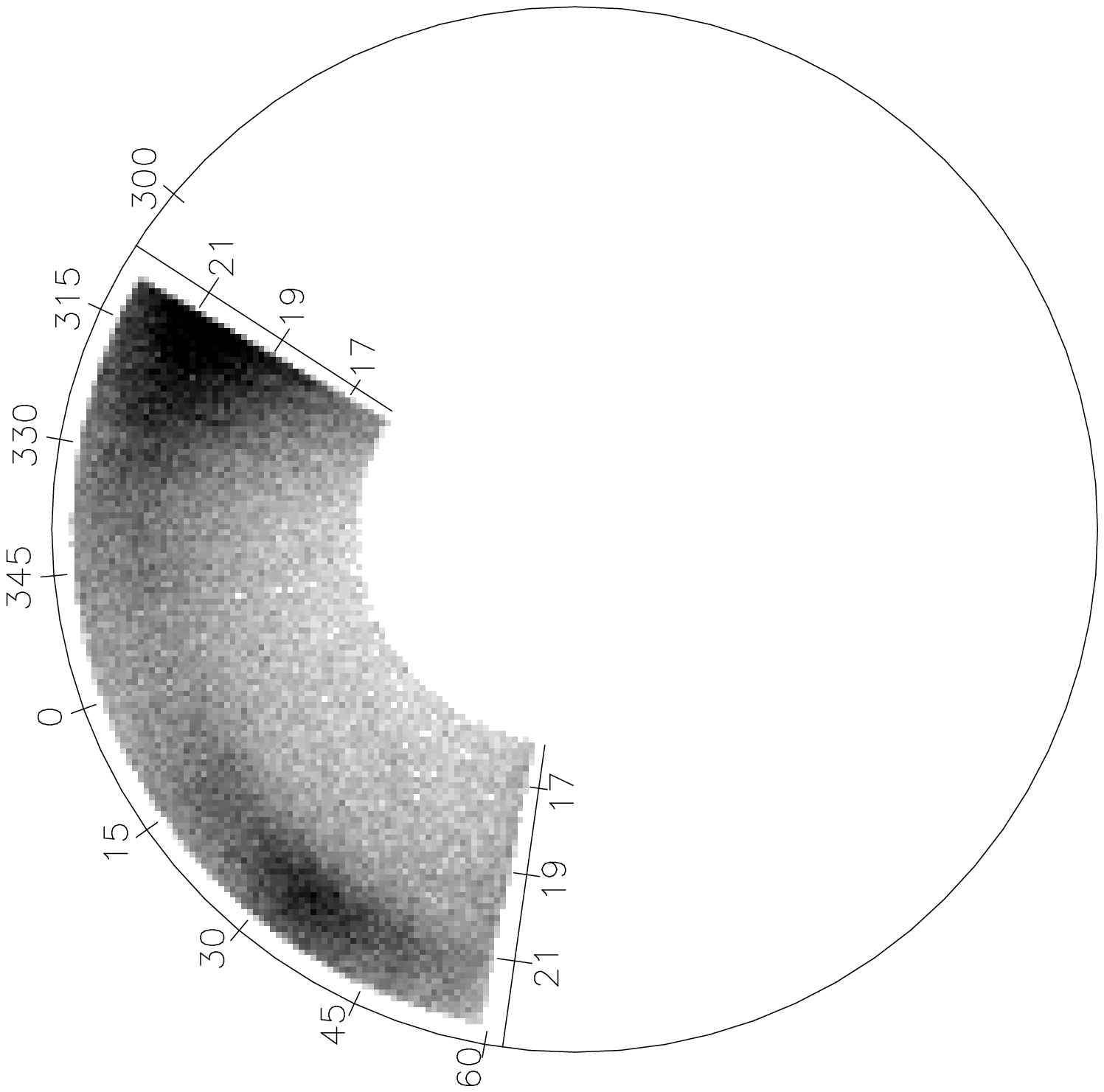}

\plotone{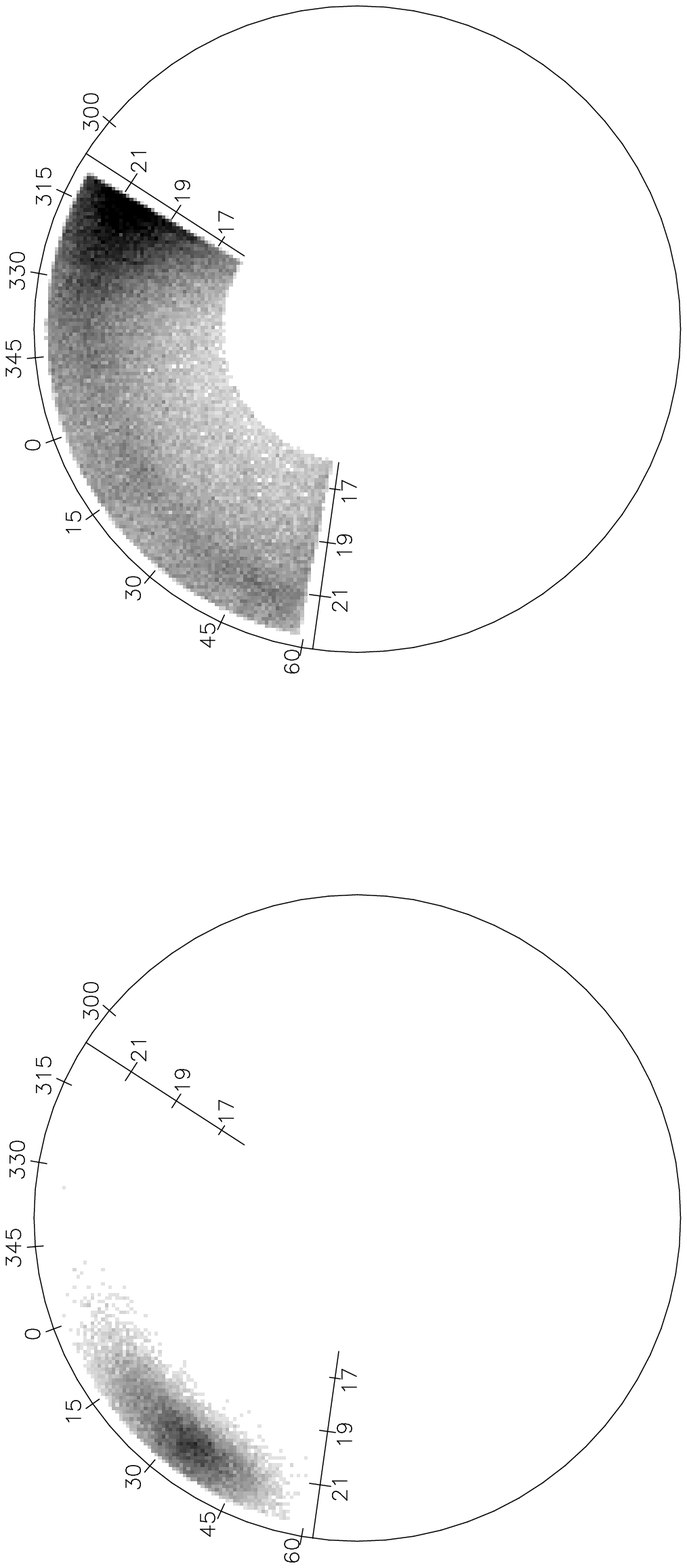}

\plotone{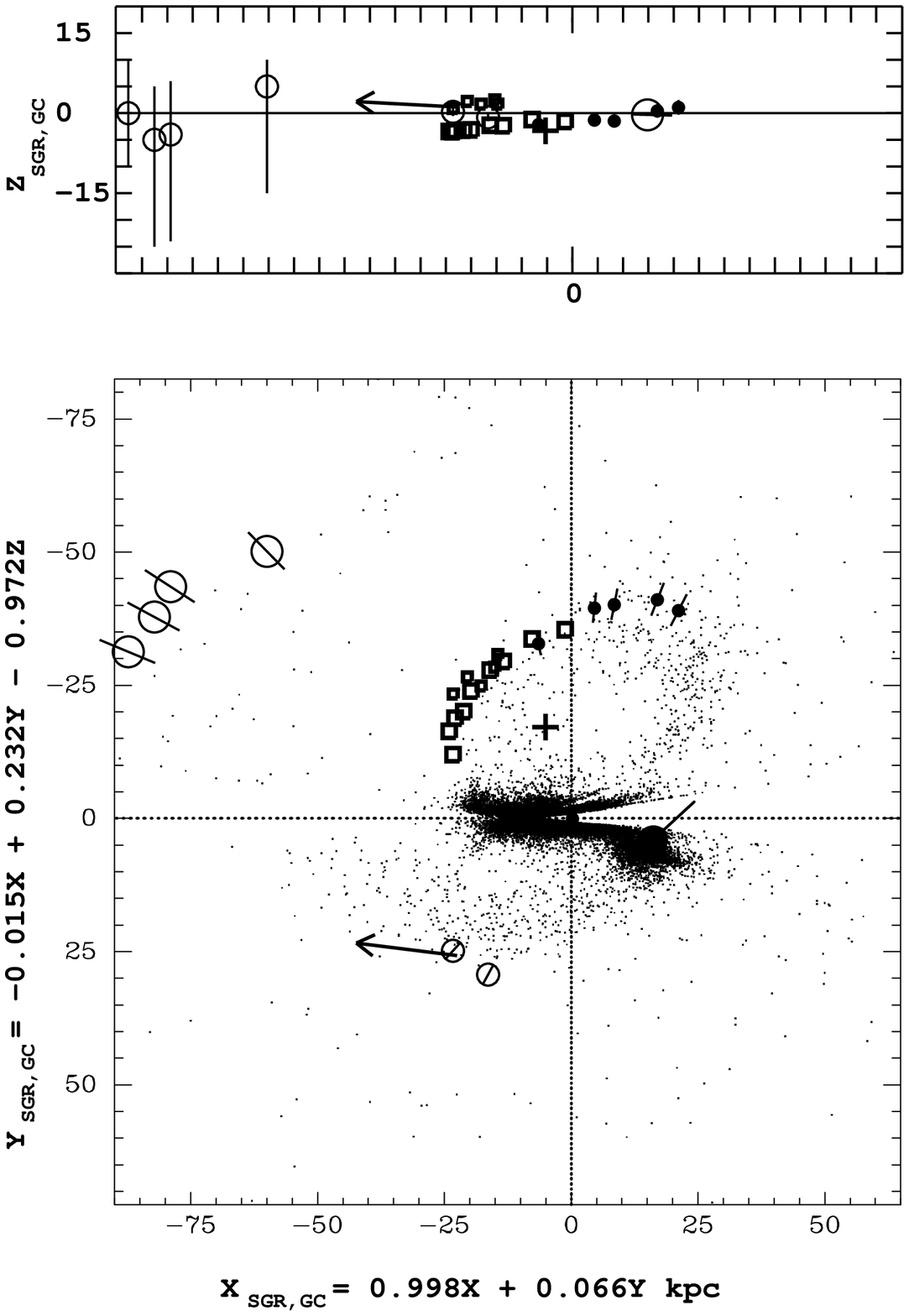}


\clearpage
\begin{thebibliography}{}
\bibitem[Abazajian et al.(2003)]{aetal03} Abazajian et al., Astron.J. 126 (2003) 2081.
\bibitem[Adelman-McCarthy et al.(2007)]{DR6} Adelman-McCarthy, J. K., et al. 2007, \apjs, submitted, arXiv:0707.3413
\bibitem[Anderson et al.(2005)]{anderson_boinc_2005} Anderson, David P., Korpela, Eric, Walton, Rom. 2005, ``High-Performance Task Distribution for Volunteer Computing", e-Science, 196-203
\bibitem[Bahcall(1986)]{b86} Bahcall, J. 1986, \araa, 24, 577
\bibitem[Bell et al.(2007)]{belletal07} Bell, Eric F., Zucker, Daniel B., Belokurov, Vasily, et al.,  2007, \apj, submitted, arXiv:0706.0004
\bibitem[Bellazzini et al.(2003)]{betal03} Bellazzini, M., Ibata, R., Ferraro, F. R., Testa, V., 2003, A\&A, 405, 577
\bibitem[Belokurov et al.(2006a)]{bootes} Belokurov, V. et al. 2006a, \apj, 647, L111
\bibitem[Belokurov et al.(2006b)]{fieldofstreams} Belokurov, V., Zucker, D. B., Evans, N.W., et al. 2006b, \apj, 642, L137
\bibitem[Belokurov et al.(2006c)]{beihw06} Belokurov, V., Evans, N.W., Irwin, M. J., Hewett, P. C., \& Wilkinson, M. I. 2006c, \apj, 637, L29
\bibitem[Belokurov et al.(2007a)]{betal07cats} Belokurov, V. et al. 2007a, \apj, 654, 897
\bibitem[Belokurov et al.(2007b)]{betal07b} Belokurov, V. et al. 2007b, \apj, 657, L89
\bibitem[Chou et al.(2007)]{cetal07} Chou, Mei-Yin, Majewski, S. R., et al. 2007, \apj, 670, 346
\bibitem[Desell, et al.(2007)]{detal07}  Desell, T., Cole, N., Magdon-Ismail, M., Newberg, H., Szymanski, B., Varela, C.  To appear December 10-13, 2007, ``Distributed and Generic Maximum Likelihood Evaluation", 3rd IEEE International Conference on e-Science and Grid Computing (eScience2007), Bangalore, India  
\bibitem[Duffau et al.(2006)]{duffau} Duffau, S., Zinn, R., Vivas, A. K., et al.  2006, \apj, 636, L97 
\bibitem[Fellhauer et al.(2006)]{fellhauer} Fellhauer, M., Belokurov, V., Evans, W. et al. 2006, \apj, 651, 167
\bibitem[Fletcher(1987)]{fletcher} Fletcher, R. 1987, Practical Methods of  Optimization, 2nd edition (New York: Wiley-Interscience) 
\bibitem[Freeman(1987)]{f87} Freeman, K. C. 1987, \araa, 25, 603
\bibitem[Freeman \& Bland-Hawthorne(2002)]{fb02} Freeman, K., and Bland-Hawthorn, J. 2002, \araa, 40, 487
\bibitem[Freese, Gondolo, \& Newberg(2005)]{fgn05} Freese, K., Gondolo, P., Newberg, H. J. 2005, Physical Review D, 71, 4
\bibitem[Fukugita et al.(1996)]{figdss96} Fukugita, M., Ichikawa,T., Gunn, J. E., Doi, M., Shimasaku, K., Schneider, D. P. 1996, \aj, 111, 1758
\bibitem[Gilmore, Wyse, and Kuijken(1989)]{gwk89} Gilmore, Wyse, Kuijken 1989, \araa, 27, 555
\bibitem[G\'{o}mez-Flechoso et al.(1999)]{gfetal99} G\'{o}mez-Flechoso, M. A., Fux, R., Martinet, L., 1999, A\&A, 347, 77
\bibitem[Grillmair(2006)]{grillmair06} Grillmair, C. J.  2006, \apj, 645, L37
\bibitem[Grillmair \& Dionatos(2006)]{gd06} Grillmair, C. J. \& Dionatos, O. 2006, \apj, 643, L17
\bibitem[Grillmair \& Dionatos(2006)]{gd06Pal5} Grillmair, C. J. \& Dionatos, O. 2006, \apj, 641, L37
\bibitem[Grillmair \& Johnson(2006)]{gj06} Grillmair, C. J. \& Johnson, R. 2006, \apj, 639, L17 
\bibitem[Gunn et al.(1998)]{getal98} Gunn, J. E. et al. 1998, \aj, 116, 3040
\bibitem[Heath(2002)]{SC} Heath, Michael, T. 2002, Scientific Computing, 2nd edition (New York, McGraw-Hill)
\bibitem[Helmi and White(2001)]{hw01} Helmi, A., White, S. D. M., 2001, \mnras, 323, 529
\bibitem[Helmi(2004)]{h04} Helmi, A., 2004, \apj, 610, L97 
\bibitem[Hernquist(1990)]{h90} Hernquist, L. 1990, \apj, 356, 359
\bibitem[Hogg et al.(2001)]{hsfg01} Hogg, D. W., Finkbeiner, D. P., Schlegel, D. J., and Gunn, J. E.  2001, \aj, 1222, 2129
\bibitem[Ibata, Gilmore, and Irwin(1994)]{igi94} Ibata, R. A., Gilmore, G., and Irwin, M. J. 1994, \nat, 370, 194
\bibitem[Ibata et al.(1997)]{ietal97} Ibata, Rodrigo A., Wyse, Rosemary F. G., et al., 1997, \aj, 113, 634
\bibitem[Ibata and Lewis(1998)]{il98} Ibata, Rodrigo A., Lewis, Geraint F., 1998, \apj, 500, 575
\bibitem[Ibata et al.(2001)]{ietal01-2} Ibata, R., Irwin, M., Lewis, G. F., and Stolte, A.  2001, \apj, 547, L133
\bibitem[Ibata, Lewis, et al.(2001)]{ietal01} Ibata, R., Lewis, G. F., et al.  2001, \apj, 551, 294
\bibitem[Irwin et al.(2007)]{ietal07} Irwin, M. J. et al. 2007, \apj, 656, L13
\bibitem[Johnston et al.(1995)]{jetal95} Johnston, K. V., Spergel, D. N., Hernquist, L., 1995, \apj, 451, 598
\bibitem[Johnston et al.(2002)]{jetal02} Johnston, K. V., Spergel, D. N., Hadyn, C., 2002, \apj, 570, 656
\bibitem[Johnston et al.(2005)]{jetal05} Johnston, K. V., Law, D. R., Majewski, S. R. 2005, \apj, 619, 800
\bibitem[Law et al.(2005)]{law05} Law, David R., Johnston, Kathryn V., and Majewkski, Steven R. 2005, \apj, 619, 807
\bibitem[Law et al.(2004)]{law04} Law, David R., et al. 2004, \pasp, 327
\bibitem[Majewski et al.(2003)]{mswo03} Majewski, S. R., Skrutskie, M. F., Weinberg, M. D., and Ostheimer, J. C.  2003, \apj, 599, 1082
\bibitem[Majewski et al.(2004)]{metal04} Majewski, S. R., Kunkel, W. E., Law, D. R., Patterson, R. J., et al. 2004, \aj, 128, 245
\bibitem[Mart\'{i}nez-Delgado et al.(2004)]{mdetal04} Mart\'{i}nez-Delgado, D., G\'{o}mez-Flechoso, M. A., Aparicio, A., and Carrera, R.  2004, \apj, 601, 242
\bibitem[Mart\'{i}nez-Delgado et al.(2007)]{mdetal07} Mart\'{i}nez-Delgado, D., Pen\~{n}arrubia, J., et al. 2007, \apj, submitted, astro-ph/0609104
\bibitem[Pen\~{n}arrubia et al.(2005)]{pmrgmnbyzg05} Pe\~{n}arrubia, J., Mart\'{i}nez-Delagdo, D., Rix, H. W., G\'{o}mez-Flechoso, M. A., Munn, J., Newberg, H., Bell, E. F., Yanny, B., Zucker, D., \& Grebel, E. K. 2005, \apj, 626, 128
\bibitem[Newberg, Yanny et al.(2002)]{nyetal02} Newberg, H., Yanny, B., et al. 2002, \apj, 569, 245 
\bibitem[Newberg, Yanny et al.(2003)]{netal03} Newberg, H., Yanny, B., et al. 2003, \apj, 596, L191
\bibitem[Newberg and Yanny(2005)]{ny05} Newberg, H. J. \& Yanny, B. 2005, A. S. P. Conf. Ser., 338, 210
\bibitem[Newberg and Yanny(2006)]{ny06} Newberg, H. \& Yanny, B. 2006, Journal of Physics: Conference Series 47, 195
\bibitem[Newberg et al.(2007)]{nyetal07} Newberg, H., Yanny, B., et al. 2007, \apj, 	arXiv:0706.3391v1
\bibitem[Odenkirchen et al.(2001)]{oetal01} Odenkirchen, M. et al. 2001, \apj, 548, L165
\bibitem[Pier et al.(2003)]{pmhhkli03} Pier, J. R., Munn, J. A., Hindsley, R. B., Hennessy, G. S., Kent, S. M., Lupton, R. H., and Ivezi\'{c}, Z. 2003, \aj, 125, 1559
\bibitem[Robin et al.(2003)]{retal03} Robin, A. C., Reyl\'{e}, C., Derri\`{e}re, S., Picaud, S., 2003, A\&A, 409, 523
\bibitem[Rocha-Pinto et al.(2004)]{rmscp04} Rocha-Pinto, H. J., Majewski, S. R., Skrutskie, M. F., Crane, J. D., and Patterson, R. J. 2004, \apj, 615, 723
\bibitem[Rockosi et al.(2002)]{retal02} Rockosi, C. M., et al. 2002, \aj, 124, 349
\bibitem[Savage et al.(2006)]{snfg06} Savage, C., Newberg, H. J., Freese, K. \& Gondolo, P. 2006, Journal of Cosmology and Astroparticle Physics, 7, 3
\bibitem[Schlegel, Finkbeiner, \& Davis(1998)]{sfd98} Schlegel, D.J., Finkbeiner, D.P., \& Davis, M. 1998, \apj, 500, 525
\bibitem[Smith et al.(2002)]{setal02} Smith, J. A. et al. 2002, \aj, 123, 2121
\bibitem[Stoughton et al.(2001)]{setal01} Stoughton, C., et al. 2001, \aj, 123, 485
\bibitem[Teuben, P.J.(1995)]{teuben} Teuben, P.J. The Stellar Dynamics Toolbox NEMO, in: Astronomical Data Analysis Software and Systems IV, ed. R. Shaw, H.E. Payne and J.J.E. Hayes. (1995), PASP Conf Series 77, p398.
\bibitem[Vivas et al.(2001)]{vetal01} Vivas, A.~K.~et al.\ 2001, \apjl, 554, L33
\bibitem[Willman et al.(2005a)]{wetal05} Willman, B. et al. 2005, \apj, 626, L85
\bibitem[Xu,Deng \& Hu(2006)]{xdh06} Xu, Y., Deng, L. C., \& Hu, J. Y. 2006, \mnras, 368, 1811
\bibitem[Yanny, Newberg et al.(2003)]{ynetal03} Yanny, B., Newberg, H. J., et al. 2003, \apj, 588, 841  
\bibitem[Yanny, Newberg et al.(2000)]{ynetal00} Yanny, B., Newberg, H. J., et al. 2000, \apj, 540, 825
\bibitem[York et al.(2000)]{yetal00} York, D. G., et al.  2000, \aj, 120, 1579
\bibitem[Zucker et al.(2006a)]{zetal06a} Zucker, D. B. et al. 2006a, \apj, 643, L103
\bibitem[Zucker et al.(2006b)]{zetal06b} Zucker, D. B. et al. 2006b, \apj, 650, L41
\end{thebibliography}
\end{document}